\documentclass[twocolumn]{aastex63}

\usepackage{xcolor,colortbl}
\usepackage{natbib}
\usepackage{mathtools}
\usepackage{longtable}
\usepackage[percent]{overpic}
\usepackage{graphicx}
\usepackage[figuresright]{rotating}
\usepackage{amsmath}
\usepackage{amsfonts}
\usepackage{graphicx}
\usepackage{rotating}
\usepackage{natbib}
\usepackage{txfonts}
\usepackage{wrapfig}
\usepackage{amssymb}
\usepackage{color}
\usepackage{xcolor}
\usepackage{ulem}

\shorttitle{Astrochemical Diagnostics of the Isolated Massive Protostar G28.20-0.05}
\shortauthors{Gorai et al.}

\begin{document}

\title{Astrochemical Diagnostics of the Isolated Massive Protostar G28.20-0.05}
\email{prasanta.astro@gmail.com}
\author[0000-0003-1602-6849]{Prasanta Gorai}
\affiliation{Department of Space, Earth \& Environment, Chalmers University of Technology, SE-412 93 Gothenburg, Sweden}
\affiliation{Department of Chemistry \& Molecular Biology , Gothenburg University, Box 462, S-40530 Gothenburg, Sweden}
\author[0000-0003-1964-970X]{Chi-Yan Law}
\affiliation{Department of Space, Earth \& Environment, Chalmers University of Technology, SE-412 93 Gothenburg, Sweden}
\affiliation{European Southern Observatory, Karl-Schwarzschild-Strasse 2, D-85748 Garching, Germany}
\author[0000-0002-3389-9142]{Jonathan C. Tan}
\affiliation{Department of Space, Earth \& Environment, Chalmers University of Technology, SE-412 93 Gothenburg, Sweden}
\affiliation{Department of Astronomy, University of Virginia, Charlottesville, VA 22904-4325, USA}
\author[0000-0001-7511-0034]{Yichen Zhang}
\affiliation{Department of Astronomy, Shanghai Jiao Tong University, 800 Dongchuan Rd., Minhang, Shanghai 200240, China}
\affiliation{Department of Astronomy, University of Virginia, Charlottesville, VA 22904-4325, USA}
\author[0000-0003-4040-4934]{Rub\'en Fedriani}
\affiliation{Instituto de Astrofísica de Andalucía, CSIC, Glorieta de la Astronomía s/n, 18008 Granada, Spain}
\affiliation{Department of Space, Earth \& Environment, Chalmers University of Technology, SE-412 93 Gothenburg, Sweden}
\author[0000-0002-6907-0926]{Kei E. I. Tanaka}
\affiliation{Department of Earth and Planetary Sciences, Tokyo Institute of Technology, Meguro, Tokyo, 152-8551, Japan}
\author[0000-0001-6551-6444]{M\'elisse  Bonfand}
\affiliation{Department of Astronomy, University of Virginia, Charlottesville, VA 22904-4325, USA}
\author[0000-0001-5551-9502]{Giuliana Cosentino}
\affiliation{Department of Space, Earth \& Environment, Chalmers University of Technology, SE-412 93 Gothenburg, Sweden}
\author[0000-0002-5065-9175]{Diego Mardones}
\affiliation{Departamento de Astronom\'ia, Universidad de Chile, Las Condes, Santiago, Chile}
\author[0000-0003-3315-5626]{Maria T. Beltr\'an}
\affiliation{INAF, Osservatorio Astrofisico di Arcetri, Largo E. Fermi 5, I-50125
Firenze, Italy}
\author[0000-0003-1649-7958]{Guido Garay}
\affiliation{Departamento de Astronom\'ia, Universidad de Chile, Las Condes, Santiago, Chile}

\begin{abstract}
We study the astrochemical diagnostics of the isolated massive protostar G28.20-0.05. We analyze data from ALMA 1.3~mm observations with resolution of 0.2\arcsec\ ($\sim$1,000 au). We detect emission from a wealth of species, including oxygen-bearing (e.g., $\rm{H_2CO}$, $\rm{CH_3OH}$, $\rm{CH_3OCH_3}$), sulfur-bearing (SO$_2$, H$_2$S) and nitrogen-bearing (e.g., HNCO, NH$_2$CHO, C$_2$H$_3$CN, C$_2$H$_5$CN) molecules. We discuss their spatial distributions, physical conditions, correlation between different species and possible chemical origins. In the central region near the protostar, we identify three hot molecular cores (HMCs). HMC1 is part of a mm continuum ring-like structure, is closest in projection to the protostar, has the highest temperature of $\sim300\:$K, and shows the most line-rich spectra. HMC2 is on the other side of the ring, has a temperature of $\sim250\:$K, and is of intermediate chemical complexity. HMC3 is further away, $\sim3,000\:$au in projection, cooler ($\sim70\:$K) and is the least line-rich. The three HMCs have similar mass surface densities ($\sim10\:{\rm{g\:cm}}^{-2}$), number densities ($n_{\rm H}\sim10^9\:{\rm{cm}}^{-3}$) and masses of a few $M_\odot$. The total gas mass in the cores and in the region out to $3,000\:$au is $\sim 25\:M_\odot$, which is comparable to that of the central protostar. Based on spatial distributions of peak line intensities as a function of excitation energy, we infer that the HMCs are externally heated by the protostar. We estimate column densities and abundances of the detected species and discuss the implications for hot core astrochemistry.

\end{abstract}

\section{Introduction}

Massive stars play a significant role in the Universe: e.g., they create heavy elements; inject energy to the interstellar medium (ISM); and regulate surrounding star formation rates. They are thus crucial for controlling the evolution of galaxies. Despite their importance, it remains unclear how massive stars form. There are various theories in contention, such as Core Accretion \citep{mcke03}, Competitive Accretion \citep[e.g.,][]{bonn01,2022MNRAS.512..216G}, and Protostellar Collisions \citep{bonn98,2020ApJ...889..178B}. These different models involve different initial conditions and evolutionary paths of the massive protostellar core, which likely translate into different chemical evolution of the gas and dust, since chemistry is highly sensitive to local physical conditions.  

Many complex physical processes, such as infall, rotation and outflows, are present, even from the earliest phases in high-mass young stellar objects (HMYSOs) and they can have a dramatic impact on astrochemical processes. A variety of species are commonly used as diagnostic tools to examine the nature of protostellar sources. For instance, CO is a good tracer of both lower-density ambient gas and protostellar outflows, C$^{18}$O is good tracer of dense gas in the envelope, and SiO is a tracer of outflow shocks. Complex organic molecules (COMs, $\ge$ 6 atoms), primarily made of carbon atoms bonded with other elements and/or other carbon atoms, are common in the heated environments of hot molecular cores (HMCs) that surround massive protostars, involving active grain surface and gas phase chemistry. Understanding the chemistry of HMYSOs is important for inferring their physical conditions and chemical inventories.

More than $300$ molecular species have been identified in the ISM and circumstellar shells [The Cologne Database for Molecular Spectroscopy (CDMS)\footnote{\url{https://cdms.astro.uni-koeln.de/classic/molecules}}]. Among them, many species are COMs that have often been
identified in HMCs \citep[e.g.,][]{herb09,jorg20}. HMCs are typically compact ($\leq$ 0.1 pc), dense ($n_{\rm H} \geq 10^{6} \:{\rm cm}^{-3}$) and hot ($\gtrsim$ 100 K) structures that are associated with luminous infra-red (IR) sources and/or with ultra-compact (UC) HII regions \citep[e.g.,][]{kurt00,vand04,cesa10}. Among the known COMs in HMCs, methanol is the most abundant and ubiquitous. Various other COMs, such as methyl formate ($\rm{CH_3OCHO}$), dimethyl ether ($\rm{CH_3OCH_3}$), acetone (CH$_3$COCH$_3$), acetaldehyde ($\rm{CH_3CHO}$), glycolaldehyde (CH$_2$OHCHO) and ethylene glycol ((CH$_2$OH)$_2$) have also been identified \citep[e.g.,][]{bell09,rivi17,paga19,jorg20}. A variety of nitrogen bearing complex species, such as formamide (NH$_2$CHO), vinyl cyanide ($\rm{C_2H_3CN}$), ethyl cyanide ($\rm{C_2H_5CN}$), methyl isocaynate (CH$_3$NCO), isopropyl cyanide (iso-$\rm{C_3H_7CN}$) have also been found \citep[e.g.,][]{bell17,bell19,colzi21}. In addition to this chemical complexity, massive star-forming regions are often associated with crowded physical environments, which can make it difficult to interpret the line profiles, chemistry, spatial distributions and chemical origins of all these molecular species. In particular, environmental effects are often invoked as the origin of chemical complexity and diversity. Thus the study of isolated massive protostars, with minimal influence from other nearby stars, is of particular importance for development and testing of astrochemical models.

G28.20-0.05 (hereafter G28.2) has been characterized as a shell-like hyper-compact HII region, and a HMC is associated with it \citep{wals03,sewi04,qin08}. The distance has been estimated as 5.7~kpc \citep{sewi08}, with this near-kinematic distance confirmed via proper motion analysis \citep{law22}. The protostar has a luminosity of $1.4\pm0.5\times 10^5\:L_\odot$ \citep{law22} \citep[see also][]{hern14,fuen20}. Previous studies \citep[e.g.,][]{soll05} suggested the presence of two components in the source: (i) an infalling equatorial gaseous torus, whose dense central region has become ionized; (ii) an extended molecular shell that is associated with a wide-angle outflow. Recent, high-resolution ALMA 1.3~mm observations indicate and that G28.2 is relatively compact, isolated and shows a ring-like structure. A detailed structural and kinematic study of this source from these data has been presented by \citet{law22}. Based on the fragmentation properties measured in the 1.3~mm continuum image, they concluded that the protostar is forming in an isolated environment, i.e., with no compact, strong 1.3~mm continuum sources identified beyond the central core of $4^{\prime\prime}$ ($0.11\:$pc) radius, and extending out over the ALMA FOV, i.e., radius of $13.5^{\prime\prime}$ ($0.37\:$pc). They identified a strong velocity gradient across the main continuum peak by investigating the H30$\alpha$ first moment map and concluded that the emission is most likely to be tracing an ionized wind that is escaping from the massive protostar. They also detected a wide, extended protostellar outflow traced by CO(2-1), which also indicates the star is still accreting. The best-fit spectral energy distribution (SED) models suggest a current protostellar mass of $43^{68}_{27}\:M_{\odot}$. From an initial inspection of the morphology of selected COMs, \citet{law22} concluded the source is chemically rich.

Previous studies detected various chemical species (e.g., CO, SO$_2$, OCS, NH$_3$, CH$_3$OH, CH$_3$CN) in G28.2 \citep[e.g.,][]{sewi08,klaa11}. However, their spatial distributions, column densities and chemical origins were not investigated because of the relatively coarse angular resolution of the data (1.3$^{\prime\prime}$ - 1.9$^{\prime\prime}$, i.e., 7,500 - 10,700 au). 
Here, we present and analyze the 1.3 mm (ALMA Band 6) line data on scales down to $0.2^{\prime\prime}$ (0.0055~pc, i.e., 1,140~au) of \citet{law22} to investigate the astrochemical inventory and implied diagnostics of G28.2.

This paper is organized as follows. In \S\ref{sec:obs-data-analysis} we describe observational details and data analysis procedures. Results are presented in \S\ref{sec:obs-results}. We discuss several aspects of our results in \S\ref{sec:discuss} and conclude in \S\ref{sec:conclusions}.

\begin{deluxetable*}{ccccccc}
\tabletypesize{\footnotesize}
\tablewidth{\textwidth}
\tablecaption{Summary of the observational parameters of the combined (compact + intermediate configuration) ALMA data, including frequency range, spectral resolution (channel spacing), beam size, and rms noise of each spectral window.\label{table:observ_sum}}
\tablehead{\colhead{SPW$^{a}$}&\colhead{Frequency}&\colhead{Channel Spacing}&\colhead{Synthesized beam}&\colhead{rms$^{b}$}\\
\colhead{}&\colhead{range (MHz)}&\colhead{(kHz)}&\colhead{($\rm{\theta_{major} \times \theta_{minor}}$)}&\colhead{mJy/beam}}
\startdata
\hline
0&232927.03 - 234927.03&15625.00&$\rm{0.20\arcsec\times0.18\arcsec}$&{0.7}\\
1& 231586.78 - 232055.53 &488.28&$\rm{0.19\arcsec\times0.18\arcsec}$&{3.2}\\
2& 230296.19 - 230764.94 &488.28&$\rm{0.18\arcsec\times0.17\arcsec}$&{5.4}\\
3& 218713.74 - 218655.14&122.07&$\rm{0.21\arcsec\times0.20\arcsec}$&{7.2}\\
4&218393.79 - 218335.20 &122.07&$\rm{0.21\arcsec\times0.20\arcsec}$&{6.2}\\
5&219513.82 - 219455.22  &122.07&$\rm{0.20\arcsec\times0.19\arcsec}$&{5.6}\\
6&220277.12 - 220218.53 &122.07&$\rm{0.21\arcsec\times0.20\arcsec}$&{6.7}\\
7&220218.53 - 216912.63 &488.28&$\rm{0.21\arcsec\times0.20\arcsec}$&{3.8}\\
8& 216685.46 - 216451.08&488.28&$\rm{0.22\arcsec\times0.21\arcsec}$&{4.2}\\
\hline
\hline
\enddata
\tablecomments{$^{a}$ SPW- Spectral window; $^{b}$rms noise level measured in the channel maps of the continuum-subtracted data
cubes.}
\end{deluxetable*}

\section{Observations and Data reduction \label{sec:obs-data-analysis}}

Observations of G28.2 were carried out with ALMA as a Cycle 3 project, covering two observational setups: Compact (C) with angular resolution $\sim$ 0.7$^{\prime\prime}$ and Intermediate (I) with an angular resolution $\sim$ 0.2$^{\prime\prime}$ (PI: Y. Zhang; 2015.1.01454.S) and via Cycle 4 project with an Extended (E) configuration with an angular resolution $\sim$ 0.03$^{\prime\prime}$ (PI: J. Tan; 2016.1.00125.S). A detailed description of the observations, including observation time, number of antennas, baseline length and maximum recoverable scale (MRS) of different configurations, is given in Table 1 of \citep{law22}. For the compact configuration, J1751+0939 (1.74~Jy) was used for band-pass and flux calibration, and J1830+0619 (0.31~Jy) was used as a phase calibrator. For the intermediate configuration, J1924-2914 (4.03~Jy) was used for band-pass and flux calibration, and J1851+0035 (0.24~Jy) was used as a phase calibrator. In each cases, the source was observed with a single pointing and the primary beam size (half power beam width) was $26.9\arcsec$. The observation was in band 6, covering a frequency range from 216 to 232~GHz. The same frequency ranges and spectral setups were used for the different configurations. 

In the E configuration, we found that most of the molecular emissions are resolved out due to very high angular resolution ($\sim$ 200 au). The resolution of the I configuration is sufficient to spatially resolve all the observed molecular species. However, the maximum recoverable scale of the intermediate configuration is only $3.4\arcsec$, so molecular emissions on larger scales are filtered out. Therefore, we combined the C configuration with I. This C+I combined data is used for the analysis in this paper. The properties of these combined data  
are given in Table \ref{table:observ_sum}. 

As part of the data reduction and analysis, raw data sets were processed through the standard ALMA calibration pipeline separately for C and I configurations by using CASA 4.5.3 and 4.7.0, respectively. All the analyses, such as continuum and spectral line analysis, were done using CASA 5.6.1 software \citep{mcmu07}. Since spectra towards G28.2 show various line emissions, it is essential to separate line and continuum information from the observed data set. To this aim, we selected line-free channels and apply first-order baseline fitting using  $\it{uvcontsub}$ task in CASA. Following this method, we separated each spectral window into two data cubes: continuum and line emission for further analysis. To construct the continuum image, we sum line-free channels. We used $\it{concat}$ task for the combination of compact and intermediate configuration continuum and line data. We used $\it{tclean}$ with `briggs' weighting and a robust factor of 0.5 for both continuum and line imaging.

After separating the continuum and line data, we performed line identification on the cleaned spectral cubes using the Centre d'Analyse Scientifique de Spectres Infrarouges et Submillimetriques (CASSIS)\footnote{\url{http://cassis.irap.omp.eu};\cite[][\url{http://adsabs.harvard.edu/abs/2015sf2a.conf..313V}]{vast15}} software, together with the Cologne Database for Molecular Spectroscopy  \citep[CDMS,][]{mull01,mull05}\footnote{\url{https://www.astro.uni-koeln.de/cdms}} and Jet Propulsion Laboratory \cite[JPL,][]{pick98}\footnote{\url{http://spec.jpl.nasa.gov}} databases.  To identify molecular transitions in the observed spectra, we first identified peaks of emission that had intensities $\geq3\sigma$, where $\sigma$ is the rms noise (see Table \ref{table:observ_sum}). The identified peaks were then cross matched against known transitions in the CDMS and JPL databases, with a focus on species already detected in the ISM. Note, the $v_{\rm LSR}$ of each hot core was determined from an average of identified strong lines from well-known species. Synthetic spectra from candidate species were also generated with a focus on conditions that are common to known hot cores (i.e., excitation temperatures up to several hundred K) and any expected multiple transitions from a given species were searched for (including allowing for potential line blending) in order to help make more secure detections.

\section{Results\label{sec:obs-results}}

\begin{deluxetable*}{cccccccccc}
\tabletypesize{\footnotesize}
\tablewidth{0pt}
\tablecaption{Positions, mass surface densities, column densities and volume densities of mm continuum peaks and identified HMCs \label{tab:cphc}}
\tablehead{\colhead{Source}&\colhead{$\alpha$(J2000)}&\colhead{$\delta$(J2000)} &\colhead{$F_\nu$}&\colhead{$\Sigma_{\rm 1.3mm}$}&\colhead{$N_{\rm H}$}&\colhead{$n_{\rm H}$}&\\
&&&(mJy)&(g cm$^{-2})$&(cm$^{-2}$)&(cm$^{-3}$)}
\startdata
Cont. peak 1&18$^h$42$^m$58.100$^s$&-4$^{\circ}$13$'$57.641$''$&138&42.16 (13.57)&$\rm{1.80\times10^{25}}$($\rm{5.80\times10^{24}}$)&$\rm{4.23\times10^{9}}$ ($\rm{1.36\times10^{9}}$)\\
Cont. peak 2&18$^h$42$^m$58.120$^s$&-4$^{\circ}$13$'$57.415$''$&93&28.41 (9.15)&$\rm{1.21\times10^{25}}$($\rm{3.91\times10^{24}}$)&$\rm{2.85\times10^{9}}$  ($\rm{9.19\times10^{8}}$)\\
HMC1 &18$^h$42$^m$58.091$^s$&-4$^{\circ}$13$'$57.841$''$&47&14.39 (4.63)&$\rm{6.15\times10^{24}}$($\rm{1.98\times10^{24}}$)&$\rm{1.44\times10^{9}}$  ($\rm{4.65\times10^{8}}$)\\
HMC2 &18$^h$42$^m$58.134$^s$&-4$^{\circ}$13$'$57.634$''$&51&15.75 (6.05) &$\rm{6.73\times10^{24}}$ ($\rm{2.58\times10^{24}}$)&$\rm{1.58\times10^{9}}$ ($\rm{6.08\times10^{8}}$)\\
HMC3 &18$^h$42$^m$58.074$^s$&-4$^{\circ}$13$'$57.058$''$&17&10.51 (15.32)&$\rm{4.49\times10^{24}}$($\rm{6.55\times10^{24}}$)&$\rm{1.05\times10^{9}}$  ($\rm{1.53\times10^{9}}$)\\
\hline
\enddata
\tablecomments{For continuum peaks 1 and 2 and HMC1 and HMC2, we remove 50\% contribution of total measured flux, $F_{\nu}$, due to an estimated contribution from free-free emission, and then use this value to estimate $\Sigma_{\rm 1.3mm}$, $N_{\rm H}$ and $n_{\rm H}$. First value in these three columns assumes $T_d=100\:$K; second value in parentheses assumes $T_d$ is at a temperature estimated from the observed molecular lines, i.e., 300~K for HMC1, 250~K for HMC2 and 70~K for HMC3. }
\end{deluxetable*}

In this section we first (in \S\ref{sec:cont-hmc}) present an overview of the continuum image and the location of peak positions of certain species. These locations define three distinct HMCs in and around the massive protostar. We next (in \S\ref{sec:spec_line}) present the full spectra extracted from the three HMCs, which are used to explore the chemical inventory of the regions, including the velocity range of the detected species. Finally (in \S\ref{sec:moment-maps}) we present integrated intensity maps of the main detected species, which more fully reveal their spatial structure. In Appendix A we also present the first moment maps of certain key species.

\subsection{Continuum image, locations of peak molecular line emissions and identification of HMCs\label{sec:cont-hmc}}

\begin{figure*}
\includegraphics[width=1.0\textwidth]{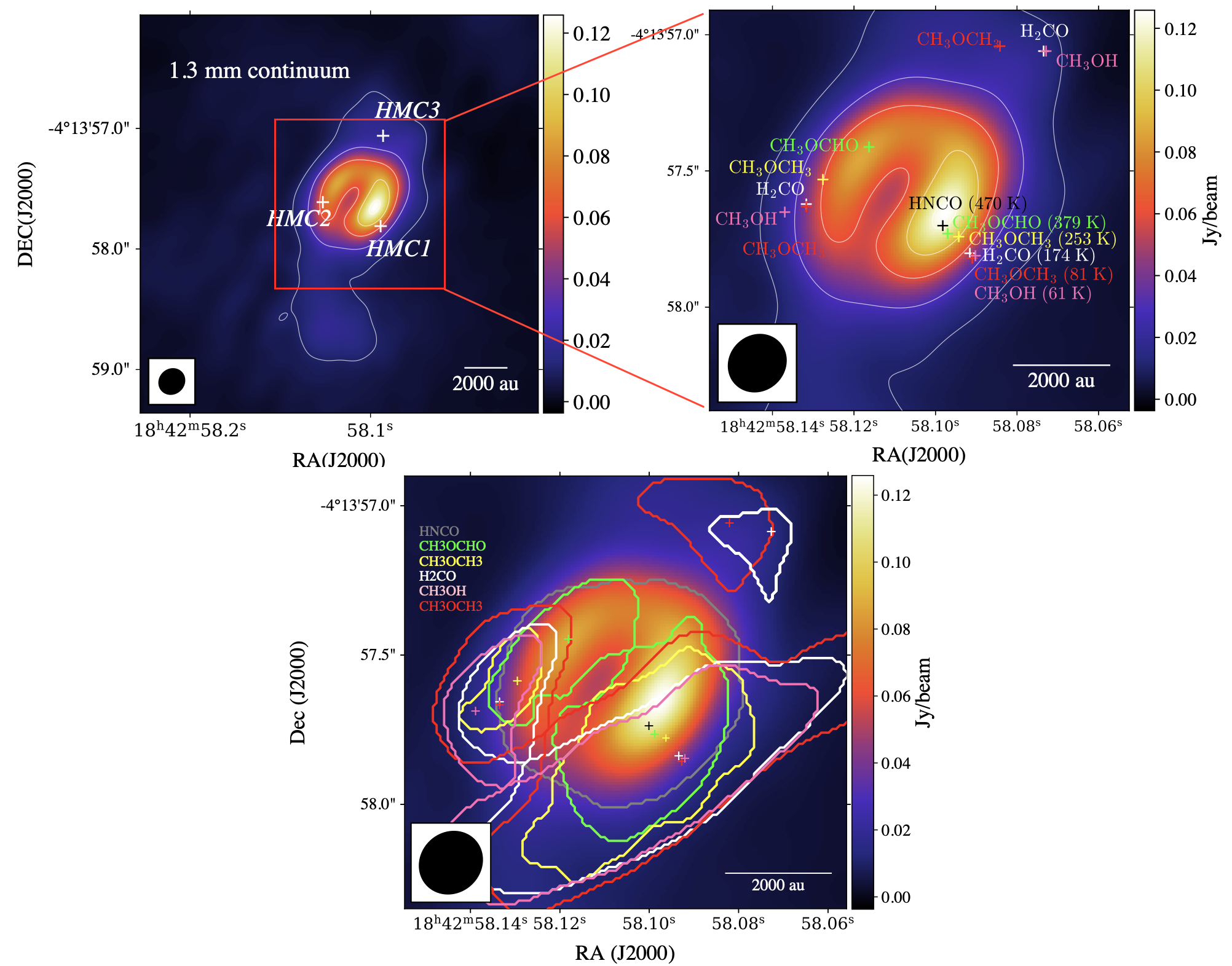}
\caption{{\it (a) Top Left Panel:} Continuum map of G28.2 obtained with ALMA at $230$ GHz. Contour levels start at 3$\sigma$ (1$\sigma$=1 mJy/beam) and steps are in  3$\sigma$. The ellipse at the bottom-left corner of the image shows the synthesized beam ($\rm{0.22\arcsec \times 0.20\arcsec}$) with PA = -47$^{\circ}$. Three white crosses represent the positions of HMC1, HMC2 and HMC3. {\it (b) Top Right Panel:} Zoom in of the central region, including locations of various molecular line emission peaks and associated upper state energies. Note the implied temperature gradients along directions pointing away from the main mm continuum peak.  {\it (c) Bottom Panel:}  Same as (b), but now showing contours of different dendrogram leaves of several identified species (see text).} 
\label{fig:cont}
\end{figure*}

Figure \ref{fig:cont} shows the continuum image of G28.2 at 1.32~mm (225.62~GHz) with the combined configuration (C+I). 
This continuum emission, including its morphology at higher resolution,
has been discussed by \citet{law22}. 
A `ring-like' structure with a radius of about 2,000 au was spatially resolved by \citet{law22} and is also discernible in the C+I data of Figure~\ref{fig:cont}.
From the mm continuum emission, after accounting for a contribution from free-free emission, \citet{law22} estimated the gas mass inside a radius of 0.5\arcsec\ to be about $30\:M_\odot$ (for an assumed dust temperature of $T_d = 100\:$K), corresponding to a mass surface density of $\Sigma_{\rm mm}=10.5\:{\rm g\:cm}^{-2}$, i.e., $N_{\rm H}=4.5\times 10^{24}\:{\rm cm}^{-2}$. This estimate is based on the assumption of optically thin dust emission, in which case a total (gas + dust) mass surface density can be derived from the 1.3 mm flux density via
\begin{equation}
\begin{aligned}
\Sigma_{\mathrm{mm}}=& 369 \frac{F_{\nu}}{\mathrm{Jy}} \frac{\left(1^{\prime \prime}\right)^{2}}{\Omega} \frac{\lambda_{1.3}^{3}}{\kappa_{\nu, 0.00638}} \\
& \times\left[\exp \left(0.111 T_{d, 100}^{-1} \lambda_{1.3}^{-1}\right)-1\right] \mathrm{g}\: \mathrm{cm}^{-2},
\end{aligned}\label{eq:dust}
\end{equation}
where $F_\nu$ is the total integrated flux over solid angle $\Omega$ and $\kappa_{\nu,0.00638} \equiv \kappa_\nu / 0.00638\:{\rm cm^2\:g^{-1}}$ is the dust absorption coefficient normalised to a fiducial value that has been derived assuming an opacity per unit dust mass of $0.899~{\rm cm^{2}\:g}^{-1}$ \citep[i.e., from the moderately coagulated thin ice mantle model of][]{osse94} and a gas-to-refractory-component-dust ratio of 141. In addition, in the above equation the dust temperature is normalized via $T_{d,100}\equiv T_d/100\:{\rm K} = 1$ and the wavelength via $\lambda_{1.3mm}\equiv \lambda/1.3\:{\rm mm}=1$.
Thus, the uncertainties in the estimate of $\Sigma_{\rm mm}$ are about a factor of two, if the temperature is uncertain with the range from 50 to 200~K.

Here, we focus on the detection of various chemical species, their spatial distributions, column densities, abundances, excitation temperatures, and possible chemical origins of individual species. In the main continuum image, HMC1, HMC2 and HMC3 are marked with white crosses, representing locations where we find a concentration of molecular line emission, as described below. 

Similar to \citet{law22}, we perform dendrogram analysis to study the different molecular structures toward G28.20-0.05 based on the integrated intensity (moment 0) maps of various species detected in spectra. We use the same dendrogram parameters as in \citep[][and references therein]{law22}, i.e.,  minvalue = $4\sigma$ (the minimum intensity considered in the analysis); mindelta = $1\sigma$ (the minimum spacing between isocontours); minpix = 0.5 beam area (the minimum number of pixels contained within a structure). 
Here $\sigma \equiv {\rm rms}_{\rm chan} \times \sqrt{N_{\rm chan}} dv$), where ${\rm rms}_{\rm chan}$ is the rms noise per channel (see Table \ref{table:observ_sum}), 
$N_{\rm chan}$ is the number of channels used to construct the moment 0 maps, and dv is the velocity resolution. For each dendrogram leaf, we identify the location of the peak intensity of its emission. We also measure the dendrogram leaf's average coordinate. 

From this analysis we notice there are three main separate groupings of the peak positions. The average of these positions are used to define the locations of the HMCs, i.e., HMC1, HMC2, and HMC3.

We define circular apertures for these HMCs of 0.15\arcsec\ radius and will present the spectra of these regions in the next section. Here we measure the continuum fluxes within these apertures and use this information to estimate the mass surface density (via Eq.~\ref{eq:dust}), and thus the total column density of H nuclei, $N_{\rm H} = \Sigma/\mu_{\rm H}=4.27 \times 10^{23} (\Sigma / {\rm g\:cm}^{-2})\:{\rm cm}^{-2}$ (i.e., where the mass per H, accounting for $n_{\rm He}=0.1n_{\rm H}$, is $\mu_{\rm H}=1.4 m_{\rm H}=2.34\times 10^{-24}\:$g). We find values of $\Sigma_{\rm 1.3mm}\sim 10\:{\rm g\:cm}^{-2}$ (all regions are within about a factor of 2 of this value, if using an estimate of $T_d$ constrained from the observed HMCs lines) (see Table \ref{tab:cphc}).

Under the assumption of spherical geometry of the core with a projected area equal to the defined circular aperture, we can also estimate the number density of H nuclei, $n_{\rm H}$. Note, if the H is mostly in molecular form, then the actual particle number density (including He) is $n_{\rm tot,mol}=0.6 n_{\rm H}$. We find values of $n_{\rm H}\sim 10^9\:{\rm cm}^{-3}$ in the hot core regions. We note that, given the variation in mm continuum flux within the apertures, we expect there is likely to be density (and temperature) variations within the defined core regions, so the above estimates should be recognized as average values in these regions.

\begin{figure*}
\begin{minipage}{0.495\textwidth}
\includegraphics[width=\textwidth]{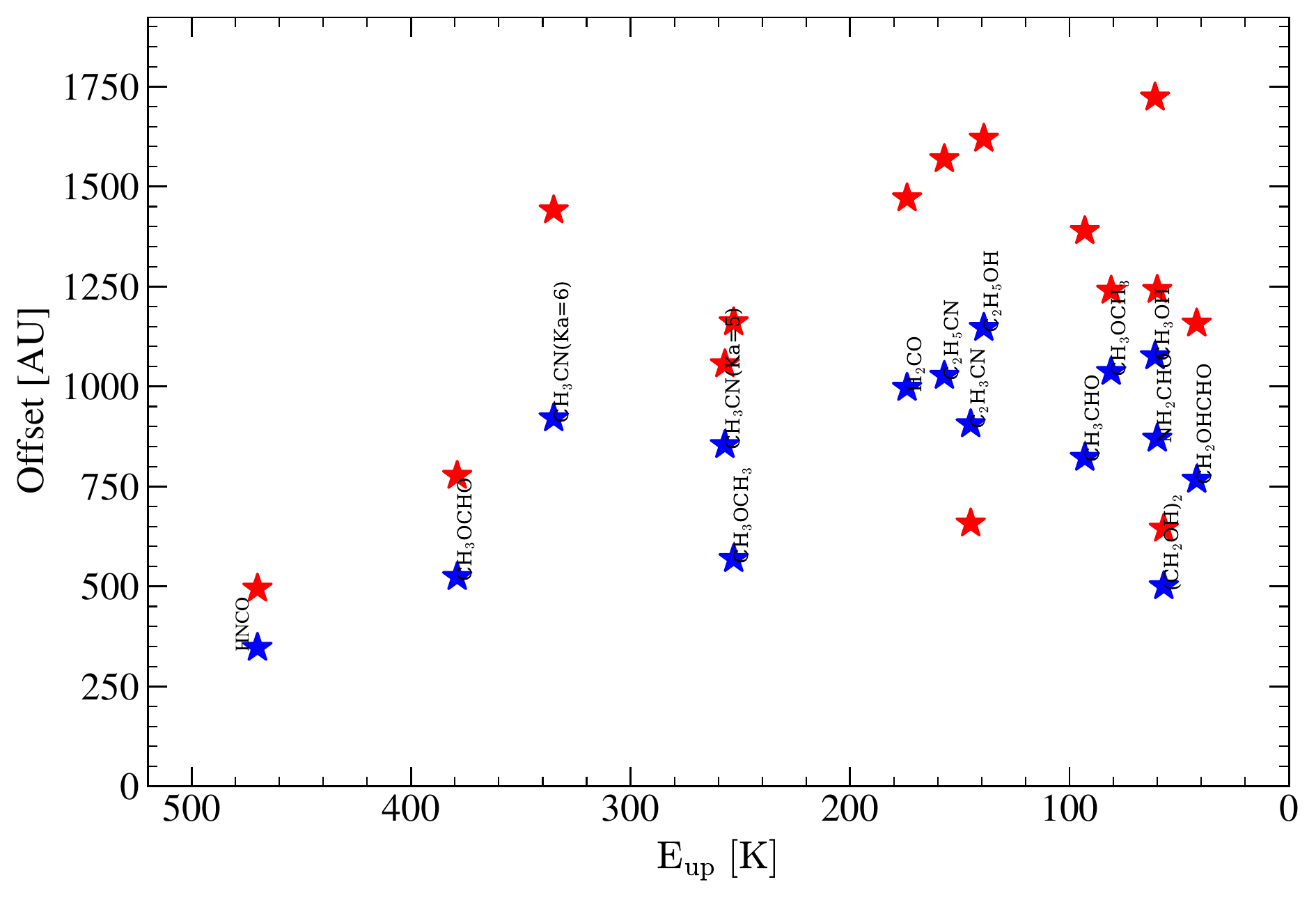}
\end{minipage}
\begin{minipage}{0.495\textwidth}
\includegraphics[width=\textwidth]{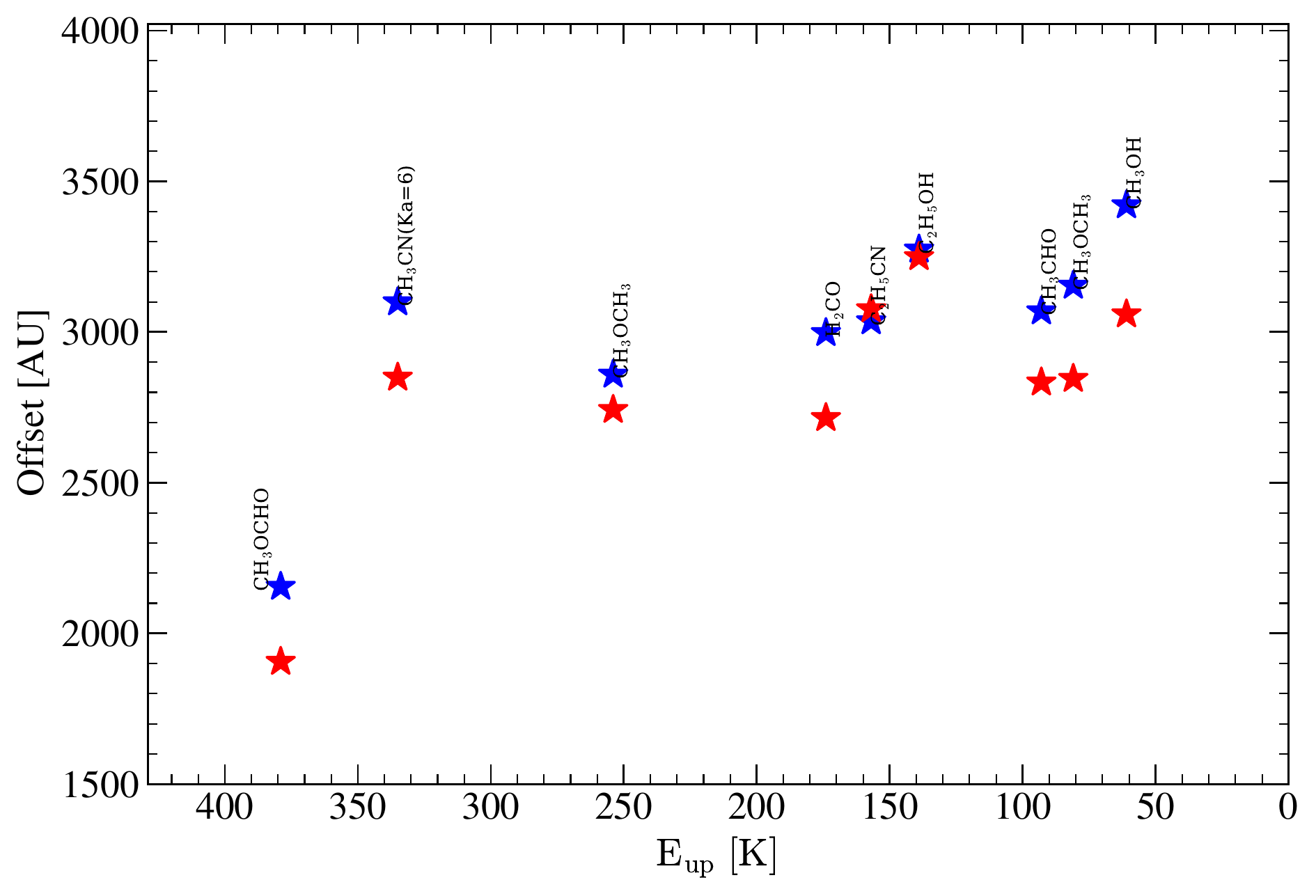}
\end{minipage}
\caption{ {\it (a) Left Panel:} Spatial offset between the molecular emission peak toward HMC1 and continuum peak 1. Blue stars are offsets of emission peaks within dendrogram leaves, while red stars are offsets of average positions of the leaves. {\it (b) Right Panel:} As (a), but now showing offsets between HMC2 line emission and continuum peak 1.}
\label{fig:offset}
\end{figure*}

In the zoom-in panels of Figure~\ref{fig:cont} we plot the peak positions for particular transitions of five species: HNCO ($E_{\rm up}=470\:$K); $\rm CH_3OCHO$ ($E_{\rm up}=379\:$K); $\rm CH_3OCH_3$ ($E_{\rm up}=254\:$K); $\rm H_2CO$ ($E_{\rm up}=174\:$K); and $\rm CH_3OH$ ($E_{\rm up}=61\:$K). All five of the species are detected in HMC1. Four of the species are detected in HMC2 (i.e., the highest energy species, HNCO, is not detected). Three of the species are detected in HMC3 (i.e., the two highest energy species are not detected). Furthermore, within HMC1 we note a spatial sequence of the peak positions, moving along a vector away from the main mm continuum peak, which is thought to be the location of the massive protostar and main heating source \citep{law22}. A similar spatial sequence is seen in HMC2 and a hint of a sequence in HMC3. In Figure \ref{fig:offset} we plot the offsets of peak emission from the main mm continuum source of a number of detected species in HMC1 and HMC2 as a function of $E_{\rm up}$. We also plot the offsets to the locations of the average dendrogram positions. For both HMC1 and HMC2 there is a trend for the highest excitation transitions to be located closest to the main mm continuum peak. We consider this to be  supporting evidence, in addition to results of \citep{law22}, that the main mm continuum peak is the location of the primary heating source in the region. Thus HMC2 and HMC3 are not likely to be self-luminous, but rather be concentrations of molecular gas heated externally by the main continuum peak. This is also largely the case for the molecular gas of HMC1, which extends to the SW of the mm continuum peak, although it does have some material overlapping (in projection) with the peak of mm continuum emission that is the likely location of the protostar. We return to a general discussion of these results in \S\ref{sec:hot-core}.

\subsection{Spectra}\label{sec:spec_line}

\begin{figure*}
\centering
\includegraphics[width=17.5cm, height=20cm]{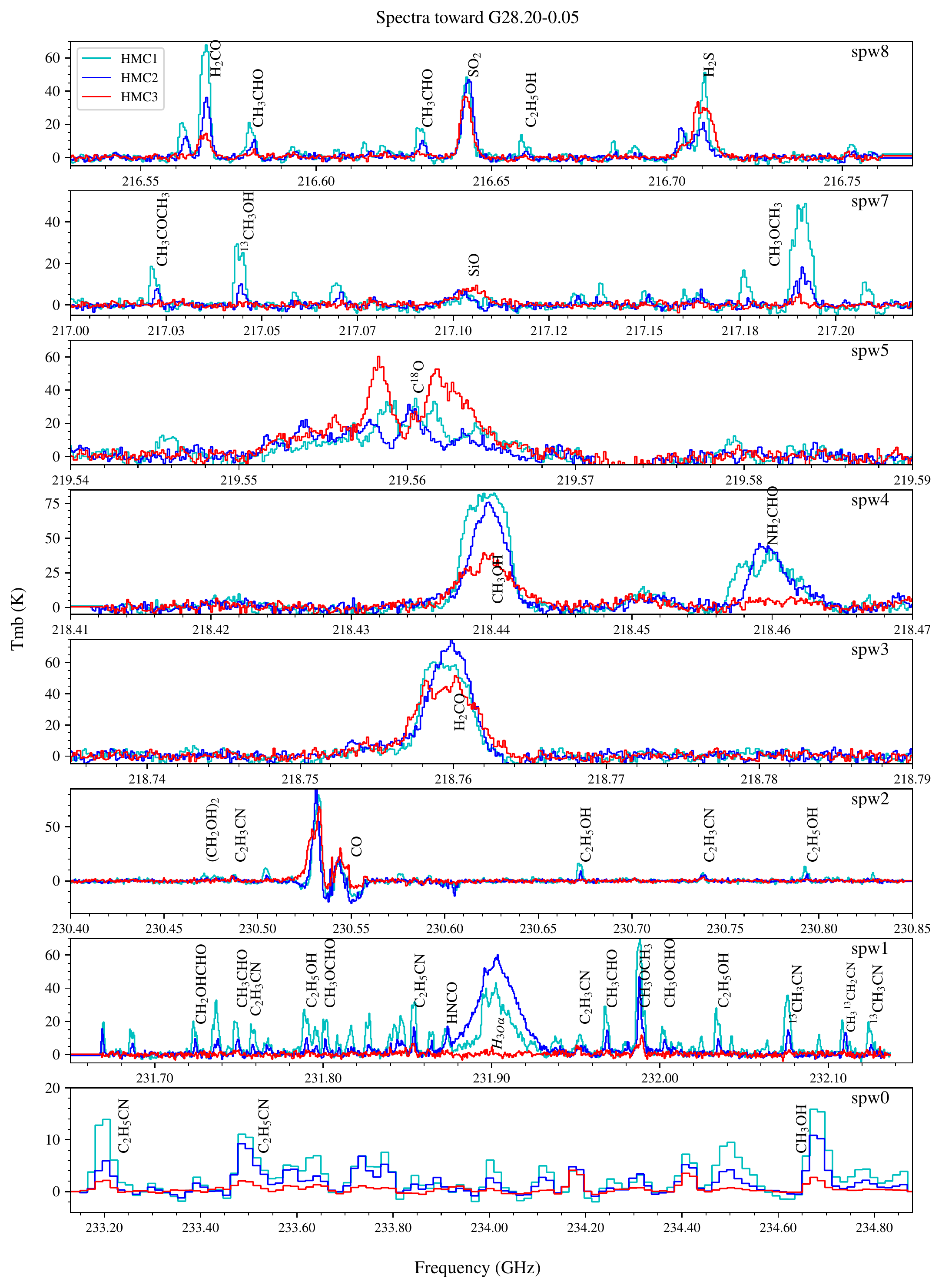}
\caption{The continuum subtracted spectra towards G28.20-0.05. All spectral windows except spw6, where we did not find any clear evidence for molecular line emission, are shown, as labelled.
Note, the displayed frequencies are in the rest frame of the source, i.e., after accounting for its $v_{\rm LSR}=95.6\:{\rm km\:s}^{-1}$.}
\label{fig:spectra}
\end{figure*}

Spectra were extracted from a circular aperture of radius of 0\farcs15 centered on the positions of HMC1, HMC2 and HMC3. 
Figure \ref{fig:spectra} shows these spectra.
Table \ref{table:LINES} summarizes all the detected species and their observed transitions. In total, 22 molecular species including isotopologues were detected, which include a wide variety of species, such as oxygen bearing ($\rm{H_2CO}$, $\rm{CH_3OCH_3}$, $\rm{CH_3CHO}$, CH$_2$OHCHO, CH$_3$CH$_2$OH), nitrogen-bearing ($^{13}$CH$_3$CN, C$_2$H$_5$CN, C$_2$H$_3$CN, and NH$_2$CHO) and sulfur-bearing (SO$_2$, H$_2$S) species (see Table \ref{table:LINES}). We note that HMC1 is the most chemically rich and its position is closest in projection to the expected location of the protostar, i.e., the main continuum peak.
HMC2, which is on the NE side of the ring, has a similar set of emission lines as HMC1, but evidence for somewhat cooler temperatures.
Finally, HMC3 is cooler and less chemically rich than HMC1 and HMC2.

Most of the observed transitions exhibit a single Gaussian-like profile. Hence, we estimate the line width (FWHM or $\Delta v$) and the peak line intensity ($I_{\rm max}$) of each transition by fitting a single Gaussian to each line profile. The parameters of all the observed transitions, such as quantum numbers ($J'_{K'_aK'_c}-J''_{K''_aK''_c}$), rest frequency ($\nu_0$), upper state energy ($E_u$), line strength (S$\mu^{2}$), FWHM, and $I_{\rm max}$ are listed in Table \ref{table:LINES}. 

The line centers of the transitions are found to be near a systematic velocity of 96.0 km s$^{-1}$ for HMC1 and HMC3, while those of HMC2 are centered at 94.5 km s$^{-1}$. The average measured line widths of almost all species in HMC1 are $\sim$5 km s$^{-1}$, except for H30$\alpha$, SiO, CO, H$_2$S and one transition of $\rm{C_2H_3CN}$, which are broader and possibly affected by the protostellar outflow. For HMC2 we find the average measured line widths of all the species are $\sim$3 km s$^{-1}$. H30$\alpha$, SiO, CO, H$_2$S, and one transition of $\rm{C_2H_3CN}$ are also broader in this region. For HMC3, the measured line widths of all species are similar to those found in HMC1.

\subsection{Spatial distribution of molecules \label{sec:moment-maps}}

\begin{figure*}
\includegraphics[width=\textwidth]{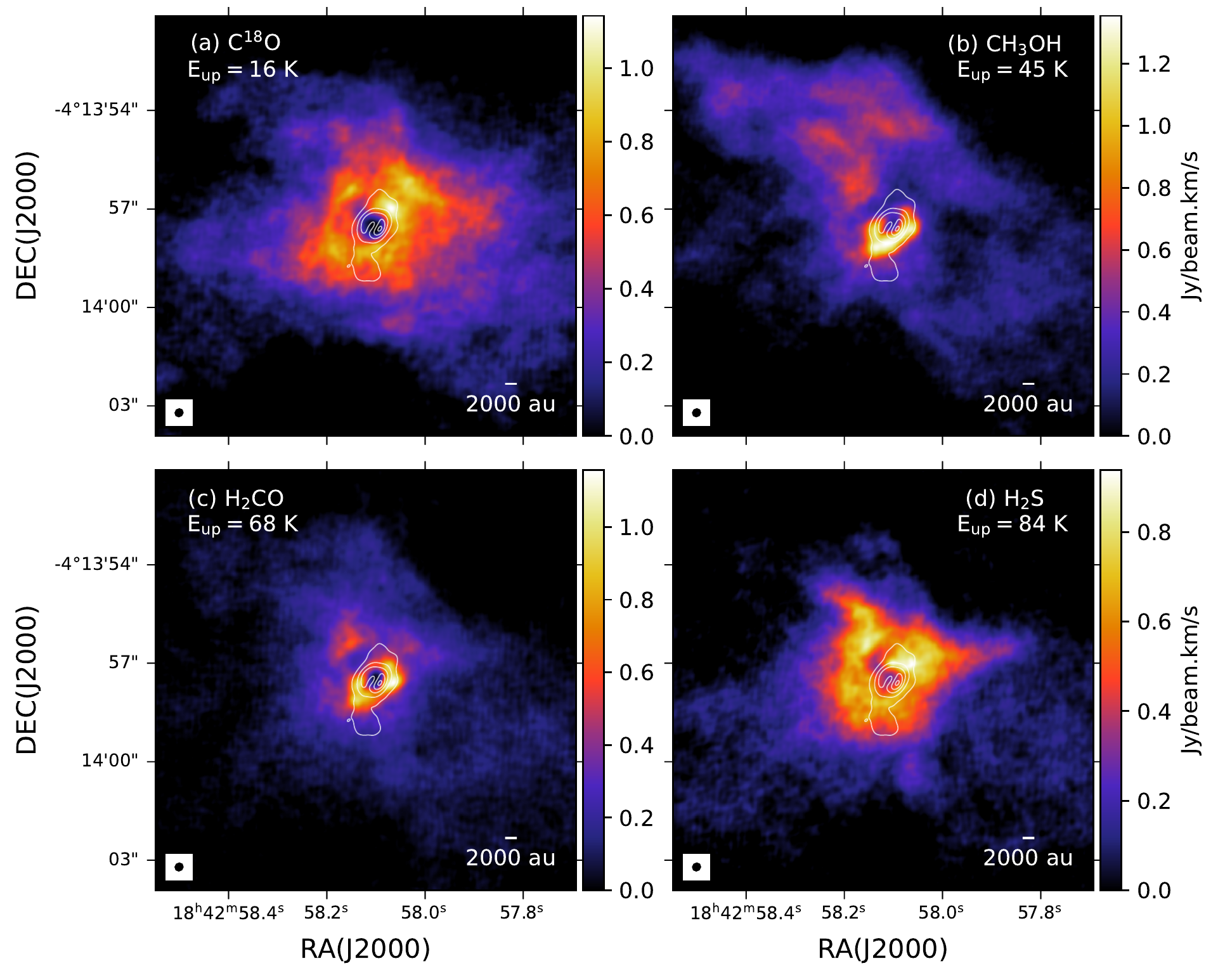}
\caption{Moment 0 maps of species showing a widespread distribution: $\rm{(a)\ C^{18}O \ (2-1; \ E_{up}=16 \ K}$), $\rm{(b)\ CH_3OH\ (E_{up}=45\ K}$), $\rm{(c)\ H_2CO\ (E_{up}=68\ K}$), and $\rm{(d)\ H_2S\ (E_{up}=84\ K}$). The white contours show 1.32~mm continuum emission (same contours as in Fig.~\ref{fig:cont}).} 
\label{fig:moment0-simple}
\end{figure*}

\begin{figure*}
\includegraphics[width=\textwidth]{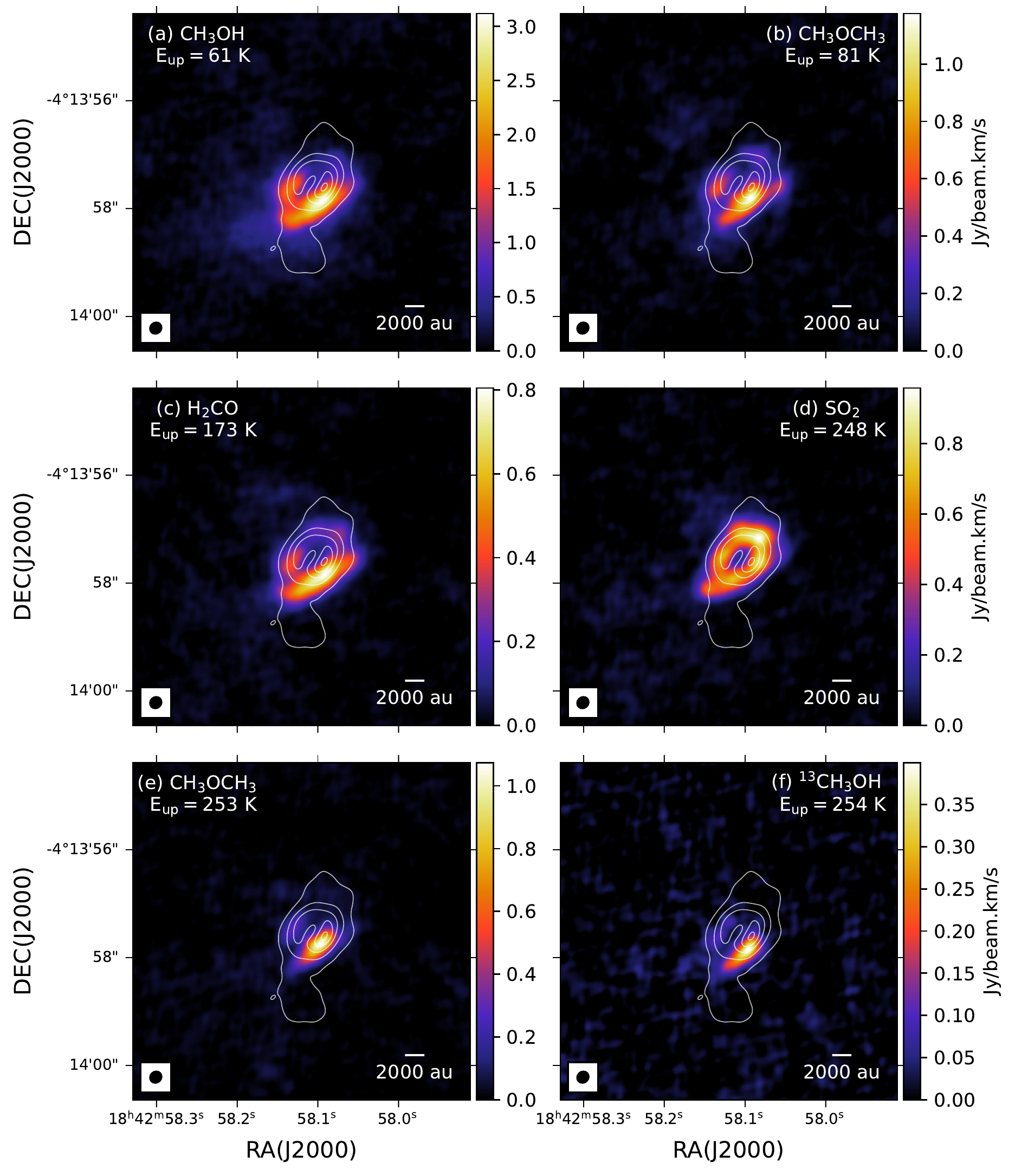}
\caption{Moment 0 maps of O-bearing species showing a concentrated distribution: $\rm{(a)\ CH_3OH \ (E_{up} = 68 \ K}$), $\rm{(b)\ CH_3OCH_3\ (E_{up} = 81\ K}$), $\rm{(c)\ H_2CO\ (E_{up} = 173\ K}$), $\rm{(d)\ SO_2\ (E_{up} = 248\ K}$), $\rm{(e)\ CH_3OCH_3\ (E_{up}=253\ K}$), and $\rm{(f)\ ^{13}CH_3OH\ (E_{up}=254\ K}$). The white contours show 1.32~mm continuum emission (same contours as in Fig.~\ref{fig:cont}).} 
\label{fig:moment0-obearing}
\end{figure*}

\begin{figure*}
\includegraphics[width=\textwidth]{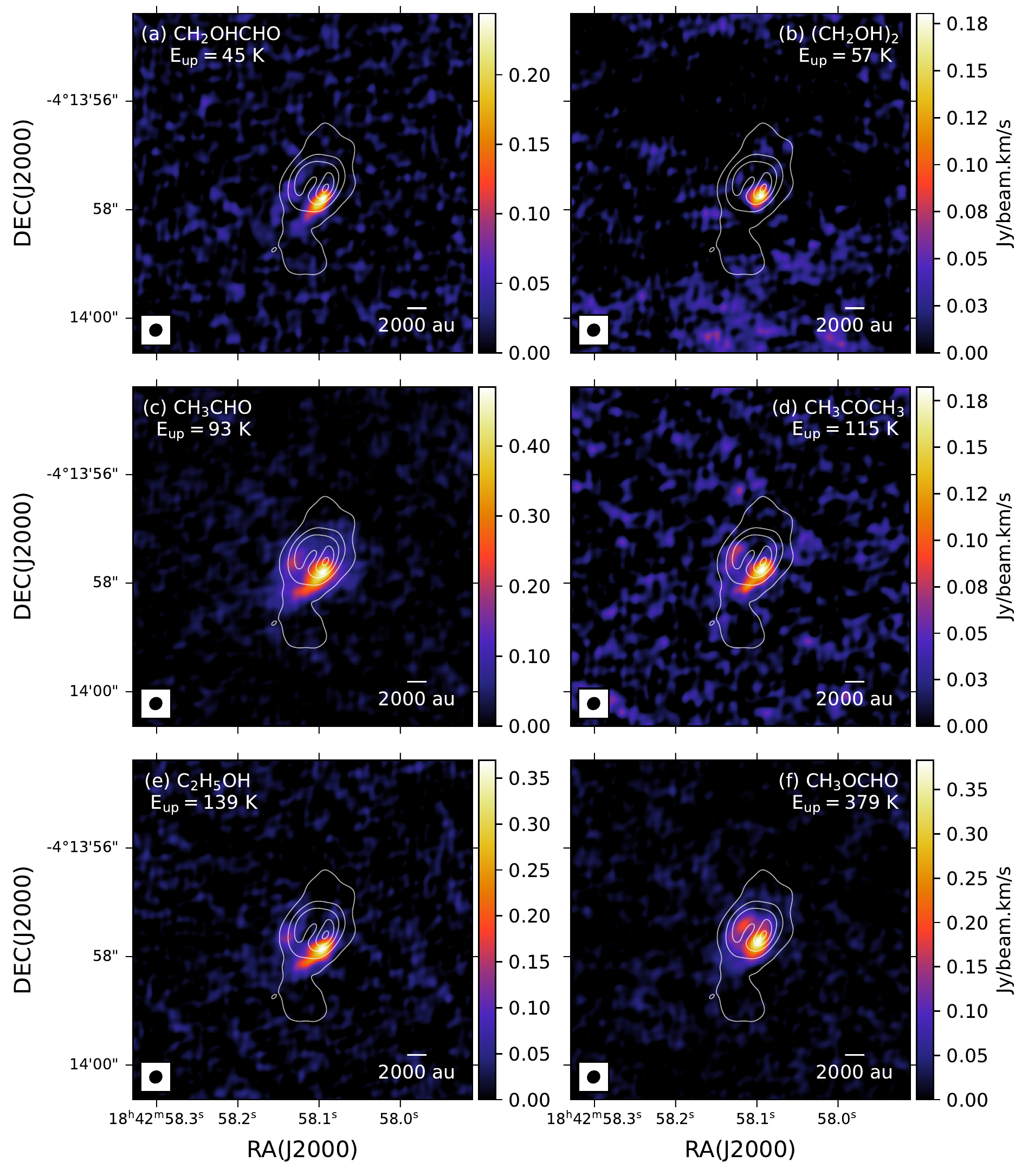}
\caption{Moment 0 maps of additional O-bearing species showing a concentrated distribution: $\rm{(a)\ CH_2OHCHO \ (E_{up} = 45 \ K}$), $\rm{(b)\ (CH_2OH)_2\ (E_{up}= 57\ K}$), $\rm{(c)\ CH_3CHO\ (E_{up}=93\ K}$), $\rm{(d)\ CH_3COCH_3\ (E_{up}=115\ K}$), $\rm{(e)\ C_2H_5OH\ (E_{up}=139\ K}$), and $\rm{(f)\ CH_3OCHO\ (E_{up}=379\ K}$). The white contours show 1.32~mm continuum emission (same contours as in Fig.~\ref{fig:cont}).} 
\label{fig:moment0-COMs}
\end{figure*}

\begin{figure*}
\includegraphics[width=\textwidth]{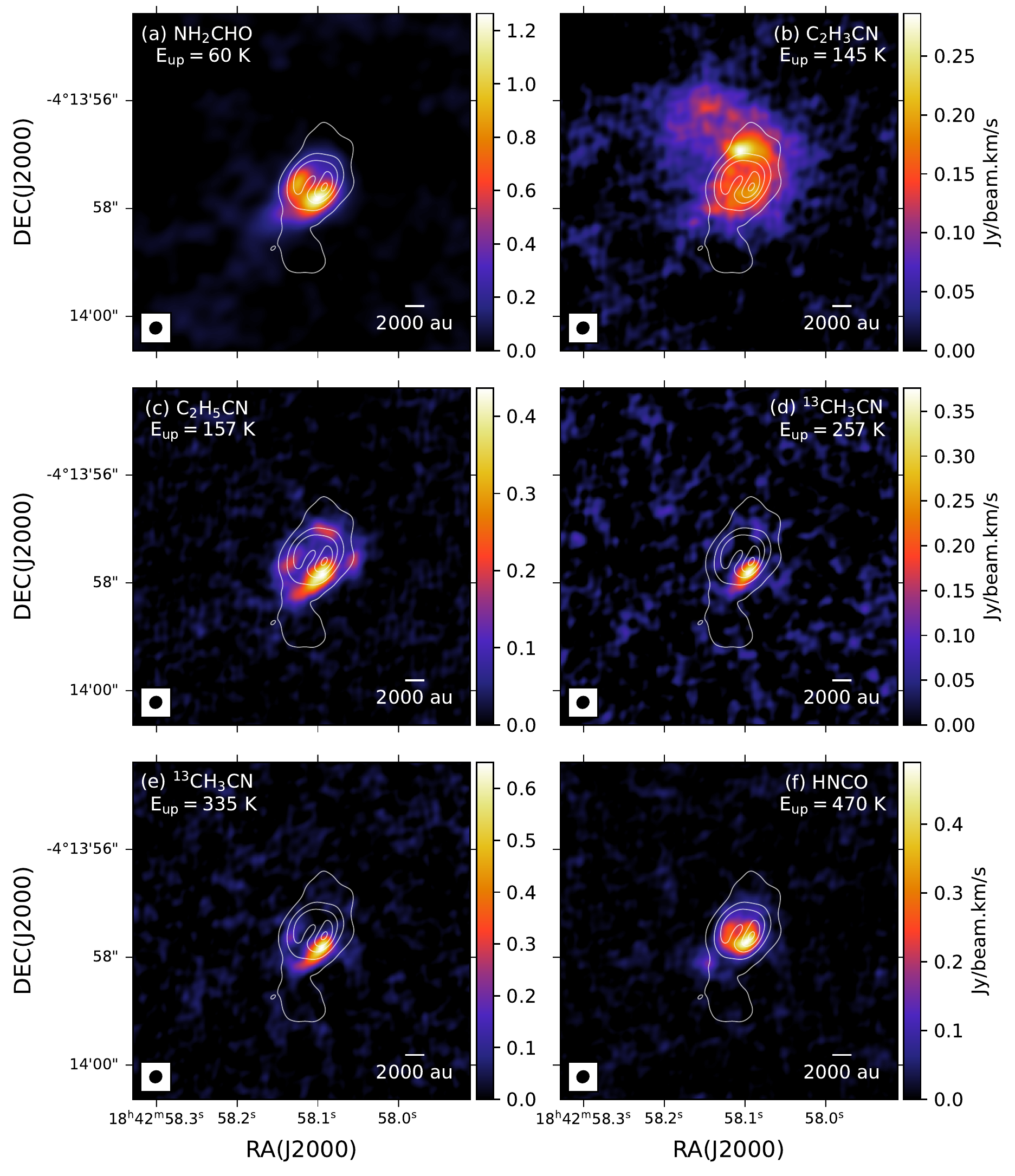}
\caption{Moment 0 maps of N-bearing species showing a concentrated distribution: $\rm{(a)\ NH_2CHO \ (E_{up} = 60 \ K}$), $\rm{(b)\ C_2H_3CN\ (E_{up}=145\ K}$), $\rm{(c)\ C_2H_5CN\ (E_{up}=157\ K}$), $\rm{(d)\ ^{13}CH_3CN\ (13_5-12_5;\ E_{up}=257\ K}$), $\rm{(e)\ ^{13}CH_3CN\ (13_6-12_6;\ E_{up}=335\ K}$), and $\rm{(f)\ HNCO\ (E_{up}=470\ K}$). The white contours show 1.32~mm continuum emission (same contours as in Fig.~\ref{fig:cont}).} 
\label{fig:moment0-nbearing}
\end{figure*}

We construct integrated intensity maps considering channels that cover the observed spectral profile of specific transitions, i.e., a velocity range of $\pm 4\:{\rm km \:s}^{-1}$ from line center. 
Figures \ref{fig:moment0-simple} - \ref{fig:moment0-nbearing} show the moment maps of all the species that are detected toward G28.2. These maps show a variety of emission peaks, including, as discussed above, varying spatial offsets from the continuum peaks.

\subsubsection{$C^{18}O$ emission}

C$^{18}$O(2-1) ($E_{\rm up}=16\:$K) is detected in a region around the massive protostar extending about 6\arcsec\ in radius, i.e., about 34,000~au or 0.17~pc. This is a region that we associate with the massive protostellar ``core'', with the $\rm C^{18}O$ tracing the infall envelope. We note there is a decrement in the C$^{18}$O(2-1) emission within the central ring-like continuum structure. While the $\rm C^{18}O$ emission exhibits substructure, this morphology is not especially correlated with the hot core positions.

\subsubsection{$H_2CO$ emission}

Two transitions of $\rm{H_2CO}$ ($J'_{K'_aK'_c}-J''_{K''_aK''_c}$ = $\rm{3_{21}-2_{20}}$; $\rm{9_{18}-9_{19}}$) are seen to be relatively strong towards G28.2 (see Table \ref{table:LINES} and Figure \ref{fig:spectra}). The FWHM of the higher excitation line ($E_{\rm up}=173\:$K) is about $0.5\:{\rm km\: s}^{-1}$ broader than that of the lower excitation line ($E_{\rm up} =68\:$K). The high excitation transition shows different peak line intensities towards the three HMCs, i.e., highest toward HMC1 and lowest toward HMC3 (see Spw 8 panel of Figure \ref{fig:spectra}). On the other hand, the low excitation transition has its highest intensity towards HMC2 and lowest toward HMC3 (see Spw 3 panel of Figure \ref{fig:spectra}). 

The moment 0 maps of these two transitions are shown in Figures \ref{fig:moment0-simple}c and \ref{fig:moment0-obearing}c. The low excitation line of $\rm{H_2CO}$ has a more concentrated spatial distribution than that of $\rm C^{18}O$. It shows enhancements near the continuum ring and a decrement within the ring. There is also a region of enhanced emission to the NE, which coincides with the direction of the near-facing outflow cavity as seen in high velocity $^{12}$CO(2-1) and SiO(5-4) emission by \citet{law22}. This could indicate a role for outflow driven shocks in liberating $\rm{H_2CO}$ into the gas phase (see \S\ref{shock} for further discussion). The emission peaks of the high excitation line of $\rm{H_2CO}$ are more spatial concentrated and are seen mainly towards HMC1 and HMC2, but also, faintly, toward HMC3.

H$_2$CO is expected to be mainly produced on grain surfaces via two successive hydrogen addition reactions with CO $\xrightarrow{\text{H}}$ HCO $\xrightarrow{\text{H}}$ H$_2$CO). It sublimes to the gas phase when conditions become warmer. 
Figure \ref{figh:moment1}a depicts the moment 1 map of the high excitation transition of $\rm{H_2CO}$. The velocity field shows complex structure, with no clear evidence for a simple gradient that might indicate rotation orthogonal to the known outflow direction (NE-SW axis).

\subsubsection{$CH_3OH$ emission}\label{sec:3_3_3}

Two transitions of $\rm{CH_3OH}$ (with $E_{\rm up}=45$ and $61\:$K) and one transition of $^{13}$CH$_3$OH (with $E_{\rm up}=254\:$K) are detected toward G28.2 (see Table \ref{table:LINES}). Figures \ref{fig:moment0-simple}b, \ref{fig:moment0-obearing}a, and \ref{fig:moment0-obearing}f show moment 0 maps of these transitions. The spatial distribution of the 61~K transition of methanol is similar to that of the high excitation line ($E_{\rm up}$= 173 K) of $\rm{H_2CO}$. Methanol is expected to be mainly produced on the surface of grains via four successive hydrogen additions with CO (CO $\xrightarrow{\text{H}}$ HCO $\xrightarrow{\text{H}}$ $\rm{H_2CO}$ $\xrightarrow{\text{H}}$ CH$_3$O $\xrightarrow{\text{H}}$ $\rm{CH_3OH}$). The morphology of methanol and formaldehyde emissions are similar, which is consistent with their similar chemical origin. Most likely they are directly evaporated from the grain mantles via various thermal and non-thermal desorption processes.

The emission of the low excitation $\rm{CH_3OH}$ transition ($E_{\rm up}$= 46 K) shows a spatial distribution that is similar to that of the low excitation line of $\rm{H_2CO}$. This indicates that these species are being liberated into the gas phase on scales of at least $\sim 10^4$~au. Extension in the outflow direction to the NE may also indicate a role for shocks in promoting this desorption
(see \S\ref{shock} for further discussion).

\subsubsection{H$_2$S emission}

One transition ($E_{\rm up}=84\:$K) of hydrogen sulfide (H$_2$S) is detected toward G28.2 (see Table \ref{table:LINES}). Figure \ref{fig:moment0-simple}d depicts the moment 0 map of H$_2$S. We find the H$_2$S emission is extended, i.e., with some similarities to the $\rm C^{18}O$ and low excitation lines of $\rm{H_2CO}$ and $\rm{CH_3OH}$, including potential enhancement in the NE outflow direction. The different spatial distribution of $\rm H_2S$ compared to $\rm SO_2$ (discussed below) indicates different chemical origins of these S-bearing species.

\subsubsection{SO$_2$ emission}

One transition of sulfur dioxide (SO$_2$) is detected ($J'_{K'_aK'_c}-J''_{K''_aK''_c}$ = 22,2,20 - 22,2,21; $E_{\rm up}$ = 253 K) toward G28.2. Figure \ref{fig:spectra} shows the observed line profile of SO$_2$. The zeroth order moment map of SO$_2$ is depicted in Figure \ref{fig:moment0-obearing}d. The emission peak of this species coincides with the ring-like continuum structure of the source as seen for H30$\alpha$ \citep{law22}. The first-order moment map is shown in Figure \ref{figh:moment1}. It reveals quite complex kinematics in the region around the continuum ring, with velocity differences of several km/s.

SO$_2$ is often observed to be associated with UC HII regions and is thought to be formed in the high-temperature gas phase through atomic sulfur or sulfur-bearing species evaporated from the grain mantles \citep{minh16b}. The similar spatial distribution of SO$_2$ and H30$\alpha$ in G28.2 may indicate that SO$_2$ is formed in the high-temperature gas.

\subsubsection{$CH_3OCH_3$ emission}

Two transitions of dimethyl ether ($\rm{CH_3OCH_3}$) are detected toward G28.2. One is of relatively low excitation ($E_{\rm up}$ = 81 K) and other is of higher excitation ($E_{\rm up}$ = 253 K) (see Table \ref{table:LINES}). The spectra of these two transitions are shown in Figure \ref{fig:spectra}. The moment 0 maps are shown in Figure \ref{fig:moment0-obearing}b and \ref{fig:moment0-obearing}e. These show brightest emission towards HMC1, fainter emission towards HMC2, and only very faint emission in the low excitation transition towards HMC3. In general, the morphology of CH$_3$OCH$_3$ is similar to those of the high $E_{\rm up}$ transitions of $\rm{CH_3OH}$, $\rm{CH_3CHO}$ and $\rm{C_2H_5OH}$, which again indicates their similar chemical origins. 

$\rm{CH_3OCH_3}$ is expected to be mainly produced on the surfaces of grain mantles via radical-radical reaction between CH$_3$ and CH$_3$O \citep[e.g.,][]{garr06,garr13}. It could also be produced in the gas phase via ion-molecule reactions in two steps starting from the reaction between protonated methanol, $\rm{CH_3OH_2^{+}}$ and $\rm{CH_3OH}$  and finally the reaction between $\rm{{(CH_3)}_2OH+}$ and NH$_3$ \citep{skou19}, with this expected to be efficient in hot core environments \citep{taqu16}.

\subsubsection{$CH_2OHCHO$ and $(CH_2OH)_{2}$ emission}

For both glyolaldehyde ($\rm{CH_2OHCHO}$) and ethylene glycol ($\rm{(CH_2OH)_{2}}$), one transition is identified toward G28.2. Figures \ref{fig:moment0-COMs}a and \ref{fig:moment0-COMs}b show the moment 0 maps of these species. Emission from both species looks similar, being compact and peaking towards HMC1. 

These two species are chemically related. $\rm{(CH_2OH)_{2}}$ could be produced via two successive hydrogen addition reactions with $\rm{HOCH_2CHO}$ \citep[e.g.,][]{rivi17,mond21}. They show similar distributions, which may indicate their similar chemical origin, i.e., sublimating together from grain mantles under similar conditions.  

\subsubsection{$CH_3CHO$ emission}

Five transitions of acetaldehyde ($\rm{CH_3CHO}$) are detected towards G28.2 (see Table \ref{table:LINES}). Among them, we have two similar sets of transitions based on their $E_{\rm up}$ values. Line widths of all these transitions are similar. The moment 0 map of one transition ($E_{\rm up}$ = 93 K) of $\rm{CH_3CHO}$ is depicted in Figure \ref{fig:moment0-COMs}c. It shows bright emission towards HMC1 and faint emission towards HMC2. We do not see any emission toward HMC3. 

This molecule is expected to be produced on grain surfaces via radical-radical reactions between HCO and CH$_3$ and between H and CH$_3$CO \citep{ruau15,rome20}. It may also be produced efficiently via gas-phase reactions \citep{vazart2020}. However, the morphological similarities with the high $E_{\rm up}$ transitions of $\rm{CH_3OH}$ and CH$_3$CHO suggest that in G28.2 it is primarily formed on grain surfaces.

\subsubsection{$C_2H_5OH$ emission}

Five transitions of ethanol ($\rm{C_2H_5OH}$) with different upper state energies are detected toward G28.2 (see Table \ref{table:LINES}). The moment 0 map of one high excitation ($E_{\rm up}$ = 244 K) transition is shown in Figure \ref{fig:moment0-COMs}e. The spatial distribution of ethanol shows that its emission is peaked towards HMC1 and similar to the spatial distribution of $\rm{CH_3CHO}$. 

The morphological correlation between $\rm{CH_3CHO}$ and $\rm{C_2H_5OH}$ emission indicates similar chemical origins. Ethanol can be produced via two successive hydrogen addition reactions with $\rm{CH_3CHO}$ and it can also be produced via various ion-neutral gas-phase and radical-radical reactions \citep[e.g.,][]{gora17a,mond21}.

\subsubsection{$CH_3COCH_3$ emission}

Two transitions with equivalent upper state energy ($E_{\rm up}$ = 155 K) of acetone ($\rm{CH_3COCH_3}$) are detected toward our target. A moment 0 map of one of these transitions is shown in Figure \ref{fig:moment0-COMs}d. It reveals that the distribution of $\rm{CH_3COCH_3}$ is similar to that of dimethyl ether, indicating some linkage of their formation routes. 

Acetone is expected to mainly form via radical-radical recombination of methyl (CH$_3$O) and acetyl radical (CH$_3$CO) radicals on the grain surfaces \citep{garr08}. Recently, \cite{sing22} studied the formation of acetone and its isomers in the laboratory with an interstellar ice analog composed of methane and acetaldehyde and proposed new branching ratios of their formation. In principle, acetone could also be produced in the gas phase via molecular radiative association, such as $\rm{CH_3^{+} +  CH_3CHO \rightarrow (CH_3)_2CHO^{+}}$, followed by dissociative recombination \citep{comb87}. However, the rates of gas-phase reactions are not expected to be as important as the surface reaction routes. 

\subsubsection{$CH_3OCHO$ emission}

Methyl formate ($\rm{CH_3OCHO}$) is commonly observed in HMCs. Two transitions of $\rm{CH_3OCHO}$ are detected toward G28.2 (see Table \ref{table:LINES}). The moment 0 map of one transition ($E_{\rm up}$ = 379 K) is shown in Figure \ref{fig:moment0-COMs}f. The distribution of $\rm{CH_3OCHO}$ emission is compact and peaks toward HMC1 and HMC2 and, as with a number of other species previously discussed, is brighter towards HMC1.

Methyl formate is expected to be formed mainly via radical-radical reactions, i.e., the reaction between CH$_3$O and CHO radicals \citep[e.g.,][]{garr06}. It may also be formed in the gas phase at low temperatures, including its precursor, methoxy (CH$_3$O) radical \citep{balu15}.

\startlongtable
\begin{deluxetable*}{p{1.7cm}p{1.8cm}cc c c c c c c c p{0.5cm}p{0.5cm}}
\tablewidth{0pt}
\tabletypesize{\footnotesize}
\tablecaption{Summary of the line parameters of observed transitions of various molecules towards G28.20-0.05.\label{table:LINES}}
\tablehead{\colhead{Species}&\colhead{$J'_{K'_aK'_c}-J''_{K''_aK''_c}$}&\colhead{Frequency}&\colhead{$E_u$ }&\colhead{$S\mu^{2}$}&\colhead{FWHM$^{1}$}&\colhead{$I_{\rm max}^{1}$}&\colhead{FWHM$^{2}$}&\colhead{$I_{\rm max}^{2}$}&\colhead{FWHM$^{3}$}&\colhead{$I_{\rm max}^{3}$}\\
\colhead{}&\colhead{}&\colhead{(GHz)}&\colhead{ (K)}&\colhead{(Debye$^{2}$)}&\colhead{(km s$^{-1}$)}&\colhead{(K)}&\colhead{(km s$^{-1}$)}&\colhead{(K)}&\colhead{(km s$^{-1}$)}&\colhead{(K)}}
\startdata
CO&2-1&230.538000&16.59&0.02&D-C&D-C&D-C&D-C&D-C&D-C\\
C$^{18}$O&2-1&219.560354&15.80&0.02&D-C&D-C&D-C&D-C&D-C&D-C\\
SiO&5-4&217.104980&31.25&47.99&5.23$\pm$0.89&5.49$\pm$0.81&8.01$\pm$0.71&6.02$\pm$0.46&11.25$\pm$0.69&7.89$\pm$0.37\\
SO$_2$&$\rm{22_{2,20}-22_{2,21}}$&216.643303&248.44&35.25&4.81$\pm$0.10&46.99$\pm$0.81&4.23$\pm$0.09&46.98$\pm$0.82&5.01$\pm$0.11&37.99$\pm$0.75\\
H$_2$S&$\rm{2_{2,0}-2_{1,1}}$&216.710436&83.98&2.06&3.47$\pm$0.16&47.36$\pm$1.83&6.94$\pm$0.49&17.65$\pm$1.01&9.19$\pm$0.48&31.43$\pm$1.30\\
$\rm{H_2CO}$&$\rm{3_{2,1}-2_{2,0}}$&218.760066&68.11&9.06&4.50$\pm$0.13&64.33$\pm$1.58&4.24$\pm$0.07&71.83$\pm$0.97&6.11$\pm$0.15&47.53$\pm$0.97\\
        &$\rm{9_{1,8}-9_{1,9}}$&216.568651&173.99&3.74&3.74$\pm$0.19&72.39$\pm$3.18&3.50$\pm$0.08&35.75$\pm$0.67&5.31$\pm$0.24&14.03$\pm$0.54\\
CH$_3$OH&$\rm{4_{2,3}-5_{1,4}}$ A, vt=0&234.683370&60.92$^{i}$&4.48&D-L&D-L&D-L&D-L&D-L&D-L\\
&$\rm{5_{4,2}-6_{3,3}}$ A, vt=0&234.698500&122.72$^{i}$&1.84&D-L&D-L&D-L&D-L&D-L&D-L\\
&$\rm{4_{-2,3}-3_{-1,2}}$ E, vt=0&218.440060&45.46$^{ii}$&13.91&4.46$\pm$0.15&88.69$\pm$2.42&3.85$\pm$0.08&73.93$\pm$1.29&5.90$\pm$0.15&33.75$\pm$0.71\\
$^{13}$CH$_3$OH&$\rm{14_{1,13}-13_{2,12}}$&217.044616&254.25&5.79&3.90$\pm$0.50&28.79$\pm$3.11&2.38$\pm$0.43&9.90$\pm$1.53&ND&ND\\
C$_2$H$_3$CN&$\rm{25_{0,25} - 24_{0,24}}$ &230.738557&145.54&1089.83&2.91$\pm$0.27&7.68$\pm$0.62&5.04$\pm$0.44&5.03$\pm$0.37&3.87$\pm$0.51&3.02$\pm$0.33\\
        &$\rm{24_{1,23} - 23_{1,22}}$&230.487936&141.25&1044.95&2.82$\pm$0.53&6.23$\pm$0.97&3.71$\pm$0.67&4.11$\pm$0.60&1.87$\pm$0.31&5.06$\pm$0.71\\
        &$\rm{24_{2,22} - 23_{2,21}}$&231.952331&146.84&1040.74&6.97$\pm$0.43&10.46$\pm$0.55&8.11$\pm$1.04&4.31$\pm$0.44&3.67$\pm$0.62&5.05$\pm$0.73\\
        &$\rm{26_{0,26} - 25_{1,25}}$&231.756777&157.04&48.35&4.57$\pm$0.60&16.17$\pm$1.73&1.86$\pm$0.26&6.29$\pm$0.76&ND&ND\\
        &$\rm{23_{5,19} - 22_{5,18}}$&218.451297&179.80&956.77&3.30$\pm$0.20&11.63$\pm$0.60&4.57$\pm$0.65&6.66$\pm$0.57&4.00$\pm$0.46&7.20$\pm$0.68\\
        &$\rm{23_{8,15} - 22_{8,14}}$&218.421801&263.81&882.78&4.98$\pm$2.93&5.08$\pm$0.85&ND&ND&ND&ND\\    
$\rm{CH_3CHO}$&$\rm{12_{3,9} - 11_{3,8}}$ &231.968385&92.62&142.33&4.14$\pm$0.36&27.14$\pm$1.99&2.57$\pm$0.18&14.41$\pm$0.82&ND&ND\\
         &$\rm{12_{3,10} - 11_{3,9}}$ &231.748719&92.51&141.14&3.94$\pm$0.30&21.37$\pm$1.37&2.98$\pm$0.23&9.12$\pm$0.61&ND&ND\\
         &$\rm{11_{1,10} - 10_{1,9}}$&216.581930&64.87&137.93&3.62$\pm$0.29&19.41$\pm$1.29&2.99$\pm$0.39&9.45$\pm$1.05&ND&ND\\
         &$\rm{11_{1,10} - 10_{1,9}}$&216.630234&64.81&137.84&3.38$\pm$0.42&19.59$\pm$2.1&3.31$\pm$0.39&9.35$\pm$0.94&ND&ND\\
C$_2$H$_5$CN&$\rm{27_{1,27} - 26_{1,26}}$&231.854212&157.73&399.21&3.74$\pm$0.31&34.10$\pm$2.40&2.34$\pm$0.18&15.69$\pm$1.01&3.18$\pm$0.25&6.99$\pm$0.47\\
        &$\rm{27_{0,27} - 26_{0,26}}$&231.990409&157.71&399.25&2.66$\pm$0.42&22.21$\pm$2.20&3.13$\pm$0.22&16.28$\pm$0.39&3.02$\pm$0.17&11.60$\pm$0.76\\
        &$\rm{26_{6,20} - 25_{6,19}}$  &233.207380&191.00&364.54&D-L&D-L&D-L&D-L&D-L&D-L\\
        &$\rm{26_{5,21} - 25_{5,20}}$  &233.498300&178.86&370.75&D-L&D-L&D-L&D-L&D-L&D-L\\            
$\rm{CH_3^{13}CH_2CN^{*}}$&$\rm{26_{6,20} - 25_{6,19}}$ &232.109851&189.24&358.46&D-B(a)&D-B(a)&ND&ND&ND&ND\\
&$\rm{26_{15,11} - 26_{15,10}}$ &231.980825&393.00&252.61&4.55$\pm$0.90&7.26$\pm$0.77&ND&ND&ND&ND\\
$\rm{CH_3OCH_3}$&$\rm{22_{4,19} - 22_{3,20}}$, AE &217.189669&253&122.88&4.46$\pm$0.62&37.88$\pm$6.41&2.04$\pm$0.23&10.59$\pm$0.91&1.98$\pm$0.97&4.77$\pm$1.30\\
            &$\rm{22_{4,19} - 22_{3,20}}$, EE&217.191400&253&327.70&3.06$\pm$1.36&23.71$\pm$5.73&1.58$\pm$0.13&16.83$\pm$1.21&1.63$\pm$1.07&3.10$\pm$1.88\\
            &$\rm{22_{4,19} - 22_{3,20}}$, AA&217.193132&253&204.82&3.40$\pm$0.69&29.84$\pm$7.85&2.93$\pm$0.29&11.20$\pm$0.78&ND&ND\\
            &$\rm{13_{0,13} - 12_{1, 12}}$, AE &231.987783&81&169.90&4.94$\pm$0.34&72.32$\pm$4.11&3.53$\pm$0.33&43.73$\pm$3.47&3.87$\pm$0.55&5.46$\pm$0.41\\
            &$\rm{13_{0, 13} - 12_{1, 12}}$, EE&231.987858&81&271.84&D-M&D-M&D-M&D-M&D-M&D-M\\
            &$\rm{13_{0, 12} - 12_{0, 12}}$, AA&231.987932&81&101.93&D-M&D-M&D-M&D-M&D-M&D-M\\
HNCO &$\rm{28_{1,28} - 29_{0,29}}$&231.873255&470.00&26.24&5.43$\pm$0.28&15.17$\pm$0.66&4.28$\pm$0.35&15.10$\pm$0.88&ND&ND\\
NH$_2$CHO&$\rm{10_{1,9} - 9_{1,8}}$&218.459213&60.81&129.35&6.06$\pm$0.31&35.62$\pm$1.41&4.01$\pm$0.15&44.20$\pm$0.98&ND&ND\\
g-$\rm{C_2H_5OH}$ &$\rm{13_{2,11} - 12_{2,10}}$&230.672554&138.62&20.28&3.72$\pm$0.34&17.06$\pm$1.35&1.56$\pm$0.11&8.98$\pm$0.54&ND&ND\\
             &$\rm{8_{4,5} - 7_{3,5}}$&216.659683&106.30&5.00&D-B(b)&D-B(b)&ND&ND&ND&ND\\
             &$\rm{6_{5,1} - 5_{4,1}}$&230.793763&104.80&5.55&3.80$\pm$0.47&11.48$\pm$1.22&1.53$\pm$0.20&6.22$\pm$0.70&ND&ND\\
t-$\rm{C_2H_5OH}$&$\rm{18_{5,14} - 18_{4,15}}$&232.034630&175.29&18.61&4.04$\pm$0.45&24.90$\pm$0.35&2.34$\pm$0.23&9.14$\pm$0.75&ND&ND\\
             &$\rm{22_{5,18} - 22_{4,19}}$&231.790000&244.54&23.10&5.67$\pm$0.53&24.58$\pm$1.75&2.36$\pm$0.08&10.14$\pm$0.27&ND&ND\\
$^{13}$$\rm{CH_3CN}$&$\rm{13_{5} - 12_{5}}$&232.125129&256.88&340.77&3.85$\pm$0.37&17.55$\pm$1.41&2.02$\pm$0.32&5.90$\pm$0.80&ND&ND\\
&$\rm{13_{6} - 12_{6}}$&232.077200&335.51&629.49&D-B(c)&D-B(c)&ND&ND&ND&ND\\
&$\rm{13_{7} - 12_{7}}$&232.002069&428.41&283.97&D-B(d)&D-B(d)&ND&ND&ND&ND\\
&$\rm{13_{8} - 12_{8}}$&231.955371&535.54&248.44&D-B(e)&D-B(e)&ND&ND&ND&ND\\
&$\rm{13_{9} - 12_{9}}$&231.881500&656.88&416.44&D-B(f)&D-B(f)&ND&ND&ND&ND\\
CH$_2$OHCHO &$\rm{8_{6,2} - 7_{5,3}}$ &231.724332&41.85&29.85&1.87$\pm$0.12&13.22$\pm$0.44&1.94$\pm$0.20&8.99$\pm$18.61&ND&ND\\
            &$\rm{8_{6,3} - 7_{5,2}}$ &231.723226&41.85&29.85&1.83$\pm$0.11&19.76$\pm$0.43&D-M&D-M&ND&ND\\
g$^{\prime}$Gg$\rm{(CH_2OH)_{2}}$&$\rm{14_{3,11} - 13_{2,12}}$&230.576100&57.10&15.35&3.38$\pm$0.70&7.24$\pm$1.25&ND&ND&ND&ND\\
g$^{\prime}$Ga$\rm{(CH_2OH)_{2}}$&$\rm{21_{4,17} - 20_{4,16}}$&230.472500&124.25&598.99&1.80$\pm$0.11&3.5$\pm$0.21&ND&ND&ND&ND\\
$\rm{CH_3OCHO}$ &$\rm{19_{11,8} - 18_{11,7}}$&232.002595&379.73&33.67&4.48$\pm$0.45&15.70$\pm$1.11&2.94$\pm$0.23&8.57$\pm$0.50&ND&ND\\
           &$\rm{19_{16,3} - 18_{16,3}}$&231.800934&469.99&14.78&2.71$\pm$0.26&21.87$\pm$0.78&D-M&D-M&ND&ND\\
           &$\rm{19_{17,2} - 18_{17,1}}$&231.801954&492.08&10.14&1.59$\pm$0.13&19.37$\pm$1.24&1.72$\pm$0.29&9.99$\pm$1.46&ND&ND\\
           &$\rm{35_{9,27} - 35_{8,28}}$&217.165400&418.42&9.70&D-B(g)&D-B(g)&D-B(g)&D-B(g)&ND&ND\\
$\rm{H_2^{13}CCO^{*}}$&$\rm{11_{1,10} - 10_{1,9}}$&217.149700&75.60&66.18&2.06$\pm$0.11&7.14$\pm$0.181&1.29$\pm$0.09&5.36$\pm$0.31&ND&ND\\
$\rm{CH_3COCH_3}$ &$\rm{19_{4,16} - 18_{3,15}}$, EE &217.022508&115.49&2042.48&3.04$\pm$0.40&176.43$\pm$0.43&1.94$\pm$0.18&8.39$\pm$0.67&ND&ND\\                          
&$\rm{19_{3,16} - 18_{4,15}}$, AA &217.070500&115.43&1276.35&4.42$\pm$0.34&10.77$\pm$0.69&2.16$\pm$0.55&5.98$\pm$1.25&ND&ND\\
\enddata
\tablecomments{$^{1}$line parameters measured towards HMC1; $^{2}$line parameters measured towards HMC2; $^{3}$line parameters measured towards HMC3; 
$^{i}$=high excitation methanol ($E_{\rm up}$=60.92 K and 122.72 K); $^{ii}$=low excitation methanol  ($E_{\rm up}$=45.46 K);  ``*''- tentative detection, ``ND'' - nondetection of transition; D-C - detection of transition, but line parameters have not been measured due to complex line profiles;  D-L - detection of transition, but in Spw 0 with low ($\sim$ 20 km s$^{-1}$) velocity resolution, so no measurement of line parameters; D-M - detection of a multiplet, but without resolution of the individual lines; D-B(a) - detection, but blended with CH$_3$CHO; D-B(b) detection, but blended with c-C$_3$H; D-B(c) - detection, but blended with C$_2$H$_5$OH; D-B(d) - detection, but blended with C$_3$HD; D-B(e) - detection, but blended with $\rm{C_2H_3CN}$; D-B(f) - detection, but blended with H30$\alpha$; D-B(g) - detection, but blended with CH$_3$COCH$_3$. RMS noise for different spectral windows varies between 0.94-2.3 K. Integrated intensity can be estimated via $\int T_{\rm mb} dv = 1.064 \times I_{\rm max} \times {\rm FWHM}$. 
}
\end{deluxetable*}

\subsubsection{HNCO emission}

Isocyanic acid (HNCO) is an important nitrogen-bearing molecule, which is the precursor of many prebiotic species (e.g., NH$_2$CHO). One transition of HNCO ($E_{\rm up}$= 470 K) is detected toward G28.2. Figure \ref{fig:moment0-nbearing}f shows the moment 0 map of HNCO. This reveals the main emission peak toward HMC1. There is also some emission toward HMC2. The morphological emission of HNCO is similar to that of other COMs species (e.g., $\rm{CH_3OH}$, $\rm{CH_3OCH_3}$) that are thought to be mostly produced on the grain surface and released to the gas phase via thermal processes.

\subsubsection{NH$_2$CHO emission\label{3_3_13}}

One transition ($E_{\rm up}$ = 61 K, see Table \ref{table:LINES}) of formamide (NH$_2$CHO) is detected toward G28.2. Figure \ref{fig:moment0-nbearing}a shows the moment 0 map of this species. The emission structure of NH$_2$CHO coincides with that of HNCO emission peaks. The derived column densities of NH$_2$CHO to HNCO (see below) is approximately 0.36.
The moment 1 map of NH$_2$CHO is depicted in Figure \ref{figh:moment1}e. The southern part is red-shifted, while the northern side is blue-shifted. The velocity separation of these regions is about 2 km s$^{-1}$. 

NH$_2$CHO has been proposed to be mainly produced on grain surfaces via two successive hydrogen addition reactions with HNCO \citep[e.g.,][]{ligt17,gora20}. NH$_2$CHO is also thought to be related to dual cyclic hydrogen addition and abstraction reactions \cite{haup19}, with this having been tested in an astrochemical model \citep{gora20}. The similar morphological structures of HNCO and N$\rm{H_2CO}$ further support the chemical link between them.

However, an alternative mechanism involves the ice-phase reaction $\rm NH_3 + CO \rightarrow NH_2CHO$ driven by ionizing radiation \citep{2017ESC.....1...50B}.
These authors \citep[][and J. Bredeh\"oft, priv. comm.]{Mues21} measure that this mechanism has a branching ratio of about 2:1 to alternatively produce $\rm HNCO + H_2$. We find a column density ratio of HNCO to NH$_2$CHO of about 1.8 in HMC1, 2.5 in HMC2 and $\sim 1/0.36 \simeq 2.8$ as an average of these and several other hot cores (see below) that is similar to this branching ratio, which lends support to this mechanism being at play. In addition, the spatial location of these species coincides with $\rm H30\alpha$ recombination line emission \citep{law22}, so the required source of ionization is likely to be present.

\subsubsection{${^{13}CH_3CN}$ emission}

Three transitions of $^{13}$C isotopologue of methyl cyanide ($\rm{^{13}CH_3CN}$) are detected toward G28.2 (see Table \ref{table:LINES}). Moment 0 maps of two different transitions are shown in Figures \ref{fig:moment0-nbearing}d and \ref{fig:moment0-nbearing}e. $\rm{CH_3CN}$ was previously identified toward G28.2 by \cite{qin08}. 
The spatial distributions of both $\rm{CH_3CN}$ and $\rm{^{13}CH_3CN}$ show similar morphologies. 

\subsubsection{$C_2H_3CN$ emission}

Vinyl cyanide ($\rm{C_2H_3CN}$) has significant importance in prebiotic chemistry, in particular by its ability to form branched chain molecules and amino acids \citep[e.g.,][]{garr17}. Five transitions of vinyl cyanide are identified toward G28.2 (see Table \ref{table:LINES}). The moment 0 map of one ($E_{\rm up}$ = 145 K) of these transitions is depicted in Figure \ref{fig:moment0-nbearing}b. It shows an emission peak in the northern dense gas blob and further extends along the axis of the outflow direction, which suggests that the formation of $\rm{C_2H_3CN}$ may be influenced by shocks in the outflowing gas. It also shows emission peaks similar to other COMs, i.e., toward HMC1. 

The $\rm{C_2H_2CN}$ radical is expected to form on the grain surface via either reaction between H and HC$_3$N or reaction between CN and C$_2$H$_2$ \citep{garr17}. However, both reactions have activation barriers, 1710 K and 1300 K, respectively. Thereafter one hydrogen addition reaction with the $\rm{C_2H_2CN}$ radical produces $\rm{C_2H_3CN}$. At about 40-50 K, $\rm{C_2H_3CN}$ could be produced in the gas phase through $\rm{C_2H_4}$ and CN \cite{garr17}. The temperatures in both HMC1 and in shocked regions are likely to be sufficient to overcome the activation barrier of this reaction. This may explain why we see peak emissions of $\rm{C_2H_3CN}$ towards both HMC1 and in the outflowing shock gas near HMC3 (see Fig. \ref{fig:moment0-nbearing}). The relative faintness of $\rm{C_2H_3CN}$ towards HMC1 compared to HMC3 is potentially due to reactions occurring efficiently at HMC1 so that much of the $\rm{C_2H_3CN}$ is already converted to $\rm{C_2H_5CN}$ due to the higher temperatures there ($\sim$300~K).

\subsubsection{$C_2H_5CN$ emission}

In the case of ethyl cyanide ($\rm{C_2H_5CN}$), four transitions are detected towards G28.2. In addition to the main species, three transitions of the $^{13}$C isotopologue of $\rm{C_2H_5CN}$ are also identified (Table \ref{table:LINES}). Moment 0 and 1 maps of C$_2$H$_5$CN are shown in Figure \ref{fig:moment0-nbearing} and Figure \ref{figh:moment1}, respectively. Figure \ref{fig:moment0-nbearing}c shows that $\rm{C_2H_5CN}$ has four distinct emission peaks, i.e., in addition to HMC1, HMC2 and HMC3 there is a fourth peak at the eastern edge of the ring.

$\rm{C_2H_5CN}$ formation is directly related to $\rm{C_2H_3CN}$ via two successive hydrogen addition reactions on the surfaces of grains \citep[e.g.,]{garr17}. The first reaction has an activation barrier of (619 - 1320 K), while the second reaction is barrier-less. As discussed in the previous section, this could be the reason why $\rm{C_2H_5CN}$ is very bright and $\rm{C_2H_3CN}$ is less bright towards HMC1 and vice versa for HMC3 and in the out-flowing gas. This could be due to the lower temperature in the northern regions away from HMC1.
We note that this species may also be formed efficiently via radical-radical reactions between $\rm{CH_3}$ and $\rm{CH_2CN}$ \citep{garr17}.

\subsection{Core Temperatures from Line Spectra}\label{sec:coreT}

Here we use the strength of certain observed lines in the spectra of HMC1, HMC2 and HMC3 to constrain their gas temperatures. We calculate synthetic spectra using local thermodynamic equilibrium (LTE) modeling with CASSIS. For such modeling, we used five adjustable parameters: (i) column density; (ii) excitation temperature ($T_{\rm ex}$); (iii) line width ($\Delta v$); (iv) velocity ($v_{\rm LSR}$); and (v) source size ($\theta_s$). Here we focus on the effects of varying $T_{\rm ex}$ to find reasonable values for the HMCs.

\begin{figure*}
\begin{minipage}{0.495\textwidth}
\includegraphics[width=\textwidth]{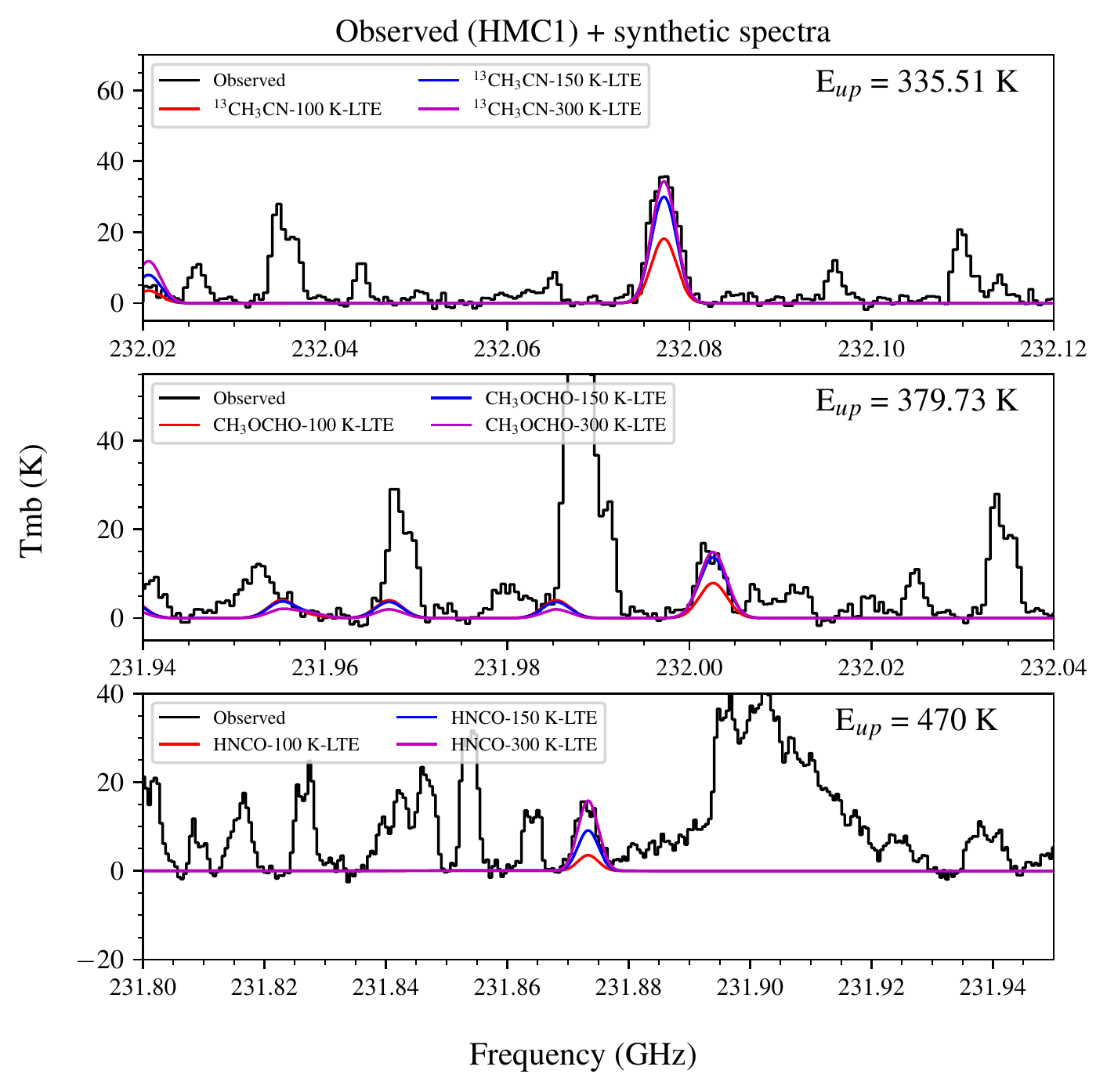}
\end{minipage}
\begin{minipage}{0.495\textwidth}
\includegraphics[width=\textwidth]{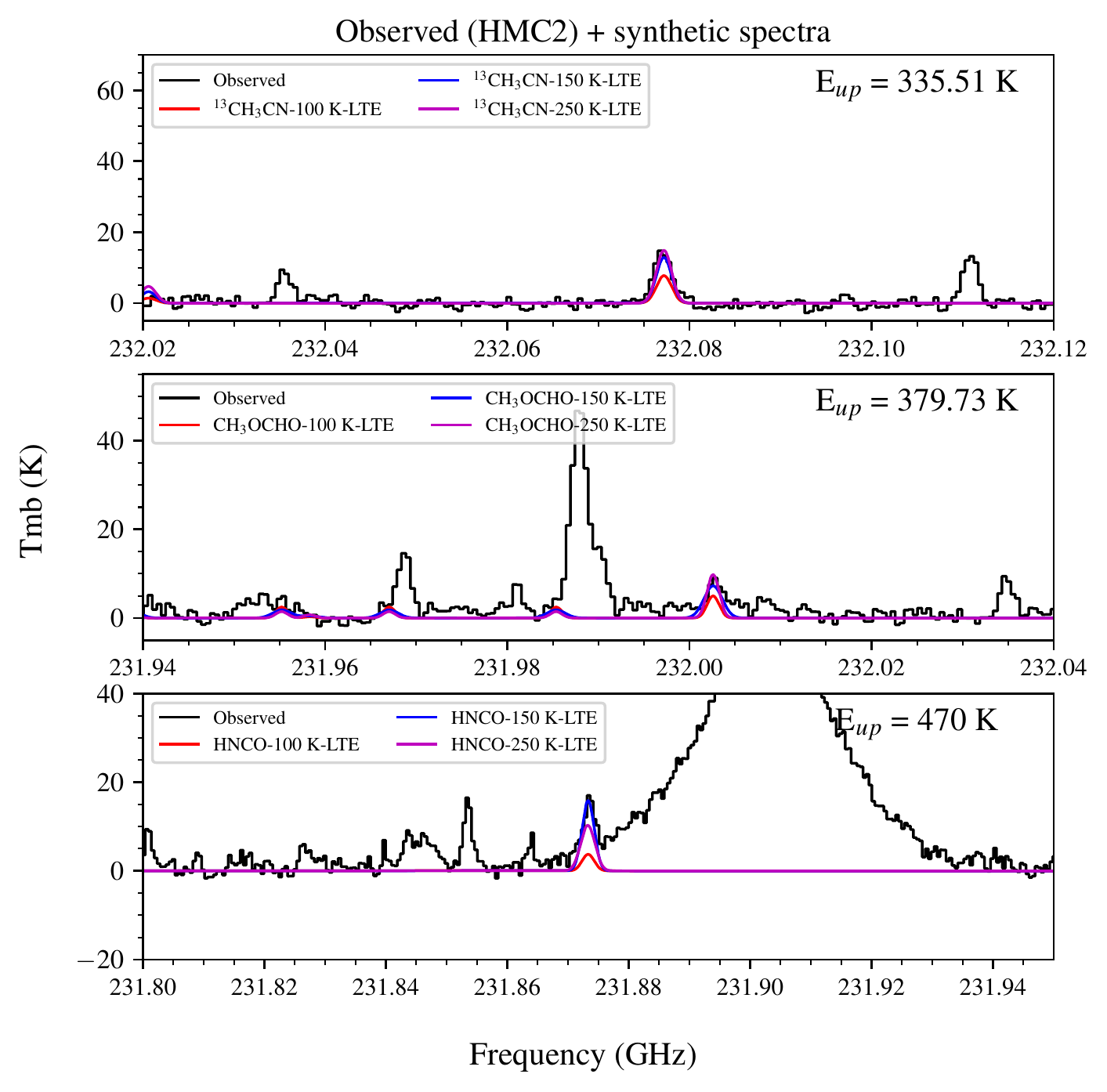}
\end{minipage}
\caption{{{\it (a) Left Panel:} Comparison between observed and synthetic spectra for HMC1, where black lines are observed spectra and red, blue, and magenta lines are synthetic spectra at 100 K, 150 K, and 300 K, respectively. {\it (b) Right Panel:} As (a), but for HMC2 and with synthetic spectra now shown for 100 K, 150, and 250 K.}}
\label{fig:lte-temp}
\end{figure*}

Figure \ref{fig:lte-temp} shows an example of synthetic spectra compared with observed spectra for HMC1 and HMC2. Here we choose three high excitation transitions of three different species: HNCO ($E_{\rm up} = 470\:$K); CH$_3$OCHO ($E_{\rm up}= 379\:$K); and $\rm{^{13}{CH_3CN}}$ ($E_{\rm up} = 335\:$K). For HMC1, we find that a fiducial temperature of $T_{\rm ex}=100\:$K is unable to reproduce the observed line profiles of these three species; instead we need a higher temperature, $\sim$ 300 K, to do so. Similarly, for HMC2, we need a temperature $\sim$ 250~K. 
However, for HMC3, which lacks detections of the above high excitation transitions, we find $T_{\rm ex}=70\:$K is a suitable temperature to reproduce the observed spectral lines.

Thus, when we evaluate column densities and abundances of species for the HMCs, we will present 
results for each core where $T_{\rm ex}$ is set to 300~K, 270~K and 70~K for HMC1, HMC2 and HMC3, respectively. We view this case as being more accurate. Physically, we also identify these fitted $T_{\rm ex}$ values as being reasonable estimates for the average gas temperatures of the HMCs. Given the high densities, we expect the dust temperature to be well coupled to the gas temperature. We have thus used these temperatures for our fiducial mass surface density estimates in \S\ref{sec:cont-hmc}. 

\subsection{Column densities of observed species}

\begin{figure*}
\includegraphics[width=\textwidth]{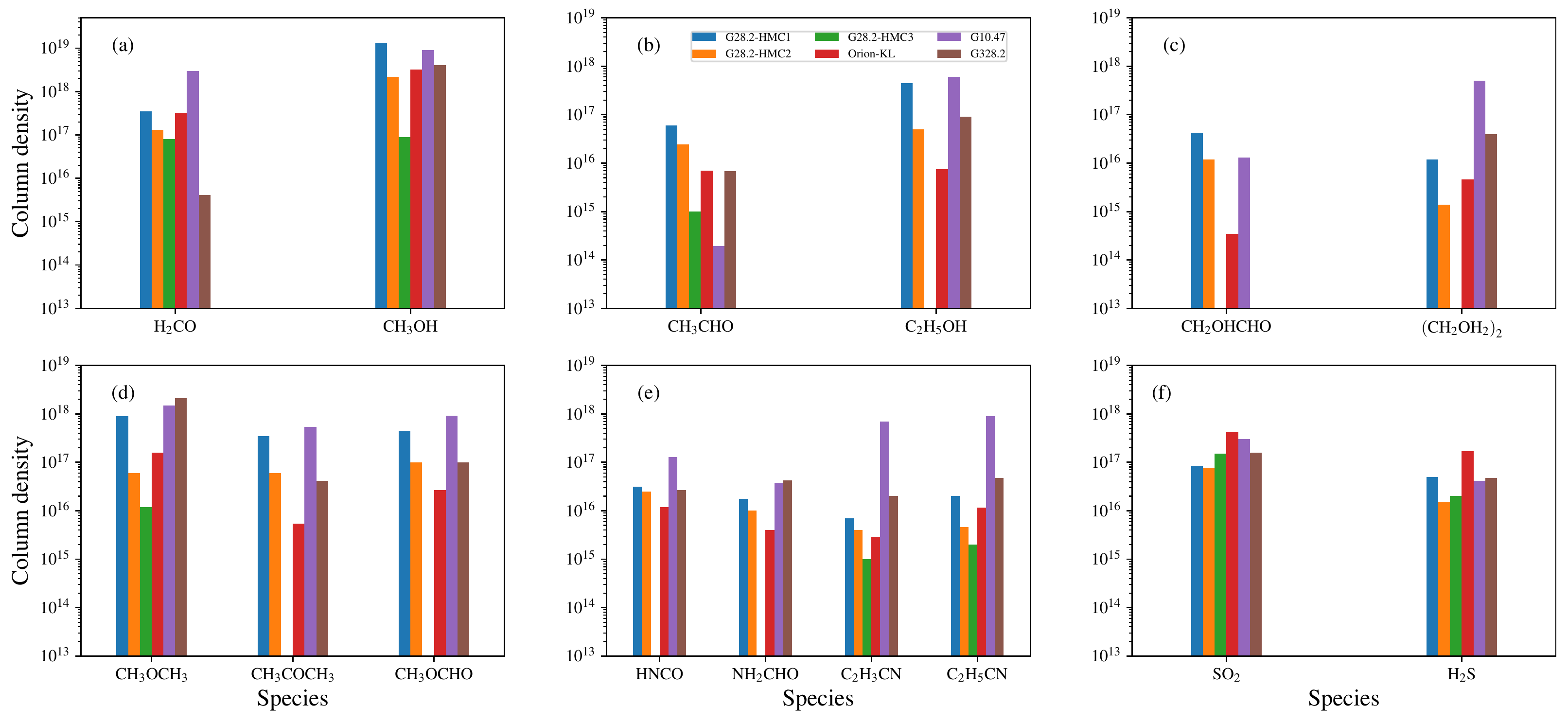}
\caption{Column densities of various observed molecules toward HMC1, HMC2 and HMC3 in G28.2 and their comparison with the equivalent results for HMCs Orion KL, G10.47 and G328.2. The first four panels (a) to (d) show oxygen-bearing molecules. Panel (e) shows nitrogen-bearing species, while panel (f) shows sulfur-bearing species. (a) Column densities of H$_2$CO and CH$_3$OH. (b) Column densities of CH$_3$CHO and C$_2$H$_5$OH. (c) Column densities of CH$_2$OHCHO and $\rm{(CH_2OH)_2}$. (d) Column densities of $\rm{CH_3OCHO, \ CH_3OCH_3\ and\ CH_3COCH_3}$. (e) Column densities of HNCO, NH$_2$CHO, $\rm{C_2H_3CN\  and\ C_2H_5CN}$. (f) Column densities of SO$_2$ and H$_2$S.}
\label{fig:column-density}
\end{figure*}

To estimate the column densities of the observed species, we use the same CASSIS modeling as in \S\ref{sec:coreT}, but now with a focus to constrain the column density of each species.
We present 
results using the individually constrained temperatures of each core from \S\ref{sec:coreT}. Note, the value of $N_{\rm H}$ has a dependence of the assumed temperature, given the results of \S\ref{sec:cont-hmc}, which thus also affects absolute abundance estimates.
The values used for line widths in the modeling are set by the observed values given in Table~\ref{table:LINES}. 
Then the column density of each species was varied to obtain a best fit to the observed spectra. Comparisons of observed and synthetic spectra for HMC1, HMC2 and HMC3 are shown in Figures~\ref{fig:synthetic1}, \ref{fig:synthetic2} and \ref{fig:synthetic3}, respectively.

The derived column density of each species ($N_{\rm X}$) and its abundance with respect to H ($N_{\rm X}/N_{\rm H}$) are listed in Table~\ref{table:mcolden} and Table~\ref{table:frac-abun} for every species detected in each of HMC1, HMC2 and HMC3. Here we also list literature results for three additional HMCs: Orion KL; G10.47; and G328.2. Figure \ref{fig:column-density} presents a visual representation of these data. Correlations between certain species that hint at chemical linkage are discussed below in \S\ref{sec:discuss}.

\section{Discussion}\label{sec:discuss}

\subsection{Nature of the hot molecular cores\label{sec:hot-core}}

We have found three HMCs, defined by the average locations of peak intensities of line emission from COMs and other species. The HMCs have properties derived from apertures of 0.15\arcsec\ radius, i.e., 850~au. These HMCs, are located in and around the mm continuum ring of the G28.2 massive protostar. HMC1 is the closest, in projection, to the massive protostar: indeed its area overlaps with this position. It is also the most chemically rich and warmest ($\sim 300\:$K). HMC2 is about 2,000~au away in projection, of intermediate chemical complexity, and slightly cooler ($\sim 250\:$K) than HMC1. HMC3 is about 3,000~au away in projection, shows the fewest lines in its spectrum, and is the coolest ($\sim 70\:$K). All the HMCs have typical $\Sigma_{\rm 1.3mm}\sim 10\:{\rm g\:cm}^{-2}$ and thus masses $\sim 1$ to $\sim 4\:M_\odot$ (note that HMC3 is inferred to be the most massive). 

By investigating the spatial offset of line emission peaks from the main mm continuum source as a function of upper state energy, we have found evidence for a gradient of temperature along vectors that point away from the massive protostar (see Figure \ref{fig:offset}). This analysis indicates that HMC1, HMC2 and HMC3 are not internally heated, i.e., all the HMCs are spatially displaced from the massive protostar. These observations also confirm that there is dense molecular gas present at all locations around the mm continuum ring.

The above results indicate the presence of significant amounts of dense, warm, clumpy molecular gas within a few thousand au of a massive protostar. The total mass of HMC1, HMC2 and HMC3 is about $7\:M_\odot$. \citet{law22} estimated a total gas mass within a 0.5\arcsec\ ($\sim3,000\:$au) radius of about $30\:M_\odot$ assuming, as we have done for HMC1 and HMC2, that half the total flux is due to free-free emission and for a constant temperature of $T_d=100\:$K. Using a value of 200~K would lower this estimate to $15\:M_\odot$. In comparison, our results that infer a temperature gradient from $\sim 300\:$K for HMC1 to $\sim 70\:$K for HMC3 imply lower gas masses in the inner region of the mm continuum ring, but larger masses in the regions just outside of it. Given the fractional areal coverage we have included in the definition of the HMCs, we estimate that the total gas mass within a region of $\sim 3,000\:$au to be about 3.7 times larger than that of the total of the cores, i.e., $\sim 25\:M_\odot$. This mass is comparable to that estimated for the current protostellar mass, i.e., $m_*\sim 40\:M_\odot$ \citep{law22}. The existence of this material on these scales is an important constraint for theoretical models and simulations of massive star formation.

The molecular gas may either be in the process of infall to join the protostar, likely via an accretion disk, or is in the process of being swept-up by the mechanical and/or radiative feedback from the star. Since there is clear evidence that the protostar is driving powerful bipolar outflows \citep[][]{law22}, we favor an interpretation that at least some of the hot core material is still in the infall phase. In this case, the observations indicate that gas is quite structured within the infall envelope. Such structure, including infalling streamers that can show significant curvature, is a natural expectation of the Turbulent Core Accretion model of massive star formation \citep{mcke03}, e.g., as seen in the simulations of \citet{2013ApJ...766...97M}. However, these authors, who simulated a $300\:M_\odot$ core with initial radius of 0.1~pc and initial magnetic field strength of 1.6~mG, did not see mass surface densities greater than $\sim 3\:{\rm g\:cm}^{-2}$ within the inner $\sim$3,000~au region, at least to the point when the protostar had grown to $16\:M_\odot$ \citep[see also Fig. 4 of][]{2014prpl.conf..149T}. Nevertheless, higher mass surface densities may arise either at later evolutionary stages or for cores that start from an initially denser state, i.e., higher surrounding clump mass surface density. 

Infall streamers to massive protostars have been previously detected and modeled, e.g., in the case of IRAS 07299-1651 \citep{2019NatAs...3..517Z}. In this case, the kinematics of an infalling streamer, as traced by CH$_3$OH, could be modeled with a simple ballistic infall solution. The kinematics of the inner region of G28.2, including its mm continuum ring, appear more complex (see, e.g., Fig. A1) and it remains uncertain whether the motions of the ring material are shaped more by infall or outflow, with the latter potentially induced by protostellar outflow, ionization or radiation pressure feedback. In particular, since H30$\alpha$ emission is detected all around the mm continuum ring \citep{law22}, then ionization feedback is likely to be playing an important role in this environment.

In terms of evidence for rotation in the molecular gas, we do not see very clear evidence for velocity gradients that are perpendicular to the outflow axis and that match the pattern of the H30$\alpha$ velocity gradient reported by \citet{law22} from analysis of the long baseline data (i.e., blueshifted to the SE and redshifted to NW of the main mm continuum peak). The H30$\alpha$ emission may be tracing the rotating surface of an ionized disk wind, which extends to larger scales than the rotating infall envelope. In general, fully rotationally supported disk structures are expected to be present on smaller scales, $\lesssim 100\:$au \citep[e.g.,][]{2013ApJ...766...97M}, which would not be resolved by the data we have analyzed, i.e., with beam of 0.2\arcsec\ equivalent to $\sim1,000\:$au.

\begin{table*}
\caption{Comparison of observed  molecular column densities between G28.2 and other sources.\label{table:mcolden}}
\centering{
\begin{tabular}{|ccccccc|}
\hline
Species & {G28.2-HMC1}& {G28.2-HMC2} & {G28.2-HMC3} &Orion KL (HMC)&G10.47+0.03& {G328.2}\\
 & {Column}& {Column} & {Column} &Column&Column&Column\\
& {density}&{density} &density&density& density & density\\
 & (cm$^{-2}$)& (cm$^{-2})$ & (cm$^{-2}$) &(cm$^{-2}$) &(cm$^{-2}$)& (cm$^{-2}$)\\
&&&&&&\\
\hline 
$\rm{N_H}$ & $\rm{1.98\times10^{24}}$& $\rm{2.58\times10^{24}}$ &$\rm{6.55\times10^{24}}$ &$\rm{4.0\times10^{24}}$&$\rm{1.0\times10^{25}}$& $\rm{3.8\times10^{24}}$
\\\hline
SO$_2$ & 8.40E+16 & 7.65E+16 & 1.50E+17 & 4.16E+17 & 3.00E+17 & 1.60E+17 \\
H$_2$S  & 5.00E+16 & 1.50E+16 & 2.00E+16 & 1.70E+17 & 4.10E+16 & 4.80E+16 \\ 
H$_2$CO & 3.50E+17 & 1.30E+17 & 8.00E+16 & 3.20E+17 & 3.00E+18 & 4.15E+15 \\ 
CH$_3$OH & 1.33E+19 & 2.21E+18 & 9.00E+16 & 3.20E+18 & 9.00E+18 & 4.10E+18 \\ 
CH$_3$CHO & 5.95E+16 & 2.45E+16 & 1.00E+15 & 6.99E+15 & 1.95E+14 & 6.83E+15 \\ 
$\rm{C_2H_5OH}$ & 4.50E+17 & 5.00E+16 & -- & 7.56E+15 & 6.00E+17 & 9.00E+16 \\ 
HNCO & 3.15E+16 & 2.50E+16 & -- & 1.19E+16 & 1.30E+17 & 2.70E+16 \\ 
NH$_2$CHO & 1.75E+16 & 1.00E+16 & -- & 4.00E+15 & 3.80E+16 & 4.20E+16 \\ 
$\rm{C_2H_3CN}$& 7.00E+15 & 4.00E+15 & 1.00E+15 & 2.91E+15 & 7.00E+17 & 2.00E+16 \\ 
$\rm{C_2H_5CN}$ & 2.00E+16 & 4.60E+15 & 2.00E+15 &1.15E+16& 9.00E+17 & 4.80E+16 \\ 
$\rm{^{13}CH_3CN}$& 6.00E+15 & 1.50E+15 & -- &-- & -- & -- \\ 
CH$_3$OCH$_3$ & 9.00E+17 & 6.00E+16 & 1.20E+16 & 1.59E+17 & 1.50E+18 & 2.10E+18 \\ 
$\rm{CH_3COCH_3}$& 3.50E+17 & 6.00E+16 & -- & 5.43E+15 & 5.40E+17 & 4.10E+16 \\ 
$\rm{CH_3OCHO}$ & 4.50E+17 & 1.00E+17 & -- & 2.69E+16 & 9.10E+17 & 1.00E+17 \\ 
$\rm{CH_2OHCHO}$ & 4.20E+16 & 1.20E+16 & -- & 3.50E+14 & 1.30E+16 & --\\ 
$\rm{(CH_2OH_2)}_2$ & 1.20E+16 & 1.40E+15 & -- & 4.60E+15 & 5.00E+17 & 4.00E+16 \\ 
\hline
\end{tabular}}
\tablecomments{
Orion KL(HMC)- \citep{feng15}- $\rm{T_{ex}}$ $\sim$ 130-200 K; resolution- 3.0$^{\arcsec}$ $\sim$ 1220 au,  \citep{mangum2005}-$\rm{T_{ex}}$ $\sim$ 250 K; resolution- 6.0$^{\arcsec}$ $\sim$ 2500 au, \citep{cern16}- $\rm{T_{ex}}$ $\sim$ 200 K; resolution- 9.0$^{\arcsec}$ $\sim$ 3700 au, \citep{luo19}- $\rm{T_{ex}}$ $\sim$ 176 K, resolution- 1.4$^{\arcsec}$  $\sim$ 580 au,  \citep{brou15}- $\rm{T_{ex}}$ $\sim$ 145 K, resolution- 1.3$^{\arcsec}$  $\sim$ 500 au, G10.47- \citep{rolf11}-$\rm{T_{ex}}$ $\sim$ 200 K; resolution- 0.35$^{\arcsec}$ $\sim$ 3700 au, \citep{gora20,mond21}-$\rm{T_{ex}}$ $\sim$ 200 K; resolution- 2.0$^{\arcsec}$ $\sim$ 21000 au,
G328.2- \citep{csen19}- $\rm{T_{ex}}$ $\sim$ 110 K; resolution- 0.23$^{\arcsec}$ $\sim$ 575 au, \citep{bous22}- $\rm{T_{ex}}$ $\sim$ 130-170 K; resolution- 14-36$^{\arcsec}$
}
\end{table*}

\begin{table*}
\caption{Comparison of observed molecular abundances w.r.t H nuclei between G28.2 and other sources.\label{table:frac-abun}}
\centering{
\begin{tabular}{|ccccccc|}
\hline
Species & {G28.2-HMC1}& {G28.2-HMC2} & {G28.2-HMC3} &Orion KL (HMC)&G10.47+0.03& {G328.2}\\
 & {Abundance}& {Abundance} & {Abundance} &Abundance&Abundance&Abundance\\
&&&&&&\\
\hline 
SO$_2$&4.24E-08 & 2.97E-08 & 2.29E-08 & 1.04E-07 & 3.00E-08 & 4.21E-08 \\ 
H$_2$S& 2.53E-08 & 5.81E-09 & 3.05E-09 & 4.25E-08 & 4.10E-09 & 1.26E-08 \\ 
H$_2$CO& 1.77E-07 & 5.04E-08 & 1.22E-08 & 8.00E-08 & 3.00E-07 & 1.09E-09 \\ 
CH$_3$OH &6.71E-06 & 8.57E-07 & 1.37E-08 & 8.00E-07 & 9.00E-07 & 1.08E-06 \\ 
CH$_3$CHO & 3.01E-08 & 9.50E-09 & 1.53E-10 & 1.75E-09 & 1.95E-11 & 1.80E-09 \\ 
$\rm{C_2H_5OH}$ &2.27E-07 & 1.94E-08 & -- & 1.89E-09 & 6.00E-08 & 2.37E-08 \\ 
HNCO   &1.59E-08 & 9.69E-09 & -- & 2.98E-09 & 1.30E-08 & 7.11E-09 \\ 
NH$_2$CHO  &8.84E-09 & 3.88E-09 & -- & 1.00E-09 & 3.80E-09 & 1.11E-08 \\ 
$\rm{C_2H_3CN}$ &3.54E-09 & 1.55E-09 & 1.53E-10 & 7.25E-10 & 7.00E-08 & 5.26E-09 \\ 
$\rm{C_2H_5CN}$ & 1.01E-08 & 1.78E-09 & 3.05E-10 & 2.88E-09 & 9.00E-08 & 1.26E-08 \\ 
$\rm{^{13}CH_3CN}$ & 3.03E-09 & 5.81E-10 & -- & -- & -- & -- \\ 
CH$_3$OCH$_3$ & 4.55E-07 & 2.33E-08 & 1.83E-09 & 3.98E-08 & 1.50E-07 & 5.53E-07 \\ 
$\rm{CH_3COCH_3}$ & 1.77E-07 & 2.33E-08 & -- & 1.36E-09 & 5.40E-08 & 1.08E-08 \\ 
$\rm{CH_3OCHO}$ & 2.27E-07 & 3.88E-08 & -- & 6.73E-09 & 9.10E-08 & 2.63E-08 \\ 
$\rm{CH_2OHCHO}$ &2.12E-08 & 4.65E-09 & -- & 8.75E-11 & 1.30E-09 & -- \\ 
$\rm{(CH_2OH_2)}_2$ & 6.06E-09 & 5.43E-10 & -- & 1.15E-09 & 5.00E-08 & 1.05E-08 \\ 
\hline
\end{tabular}}
\tablecomments{Orion KL(HMC)- \citep{feng15}- $\rm{T_{ex}}$ $\sim$ 130-200 K; resolution- 3.0$^{\arcsec}$ $\sim$ 1220 au,  \citep{mangum2005}, \citep{cern16}- $\rm{T_{ex}}$ $\sim$ 200 K; resolution- 9.0$^{\arcsec}$ $\sim$ 3700 au, \citep{luo19}- $\rm{T_{ex}}$ $\sim$ 176 K, resolution- 1.4$^{\arcsec}$  $\sim$ 580 au,  \citep{brou15}- $\rm{T_{ex}}$ $\sim$ 145 K, resolution- 1.3$^{\arcsec}$  $\sim$ 500 au, G10.47- \citep{rolf11}-$\rm{T_{ex}}$ $\sim$ 200 K; resolution- 0.35$^{\arcsec}$ $\sim$ 3700 au, \citep{gora20,mond21}-$\rm{T_{ex}}$ $\sim$ 200 K; resolution- 2.0$^{\arcsec}$ $\sim$ 21000 au,
G328.2- \citep{csen19}- $\rm{T_{ex}}$ $\sim$ 110 K; resolution- 0.23$^{\arcsec}$ $\sim$ 575 au, \citep{bous22}- $\rm{T_{ex}}$ $\sim$ 130-170 K; resolution- 14-36$^{\arcsec}$
}
\end{table*}

\subsection{Comparison of column densities of different species}\label{sec:column}

\begin{figure*}
\includegraphics[width=\textwidth]{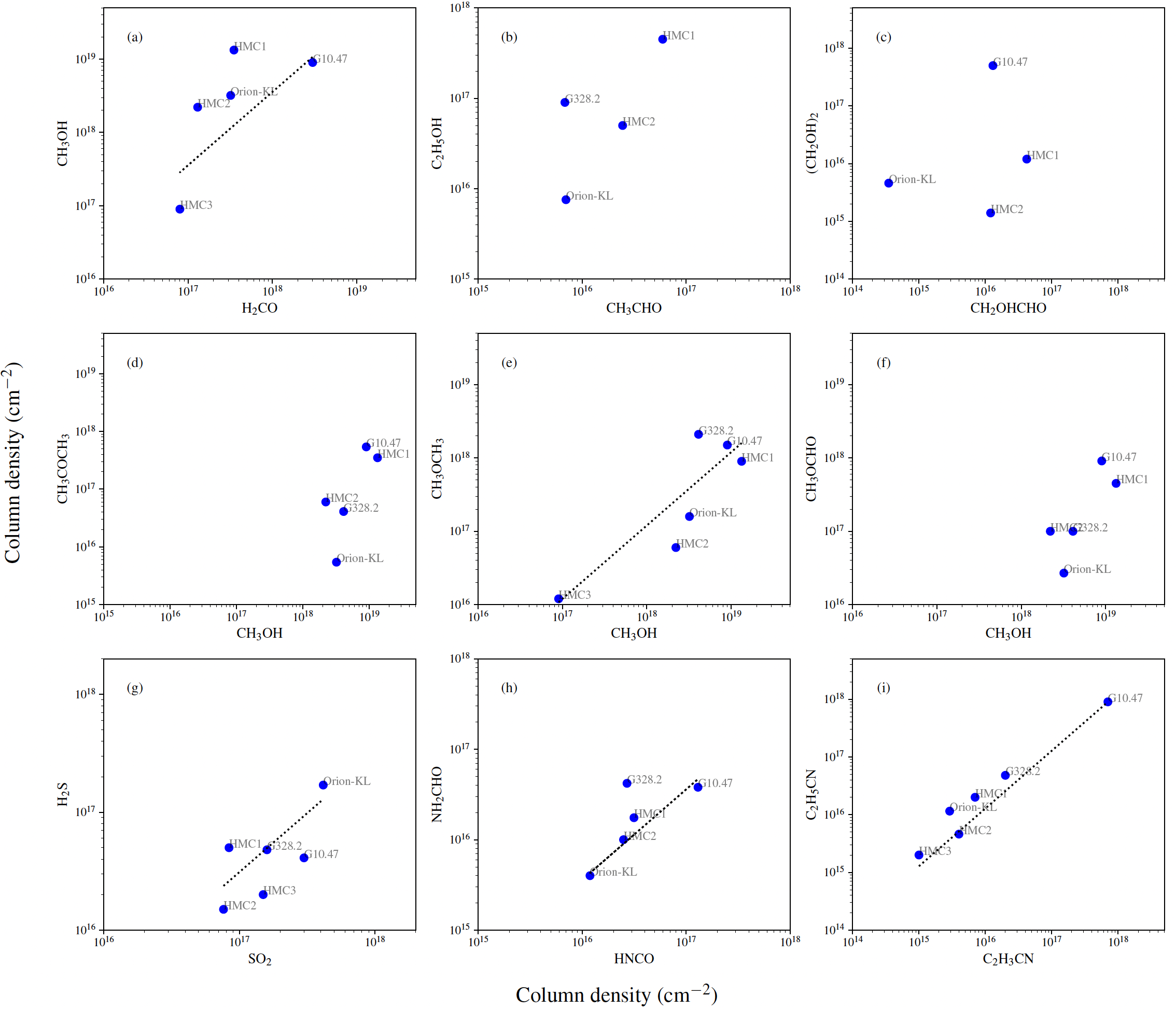}
\caption{Correlations between column densities of different pairs of molecules in the G28.2 HMCs and three other HMCs, providing information on abundance ratios. (a) CH$_3$OH versus H$_2$CO, with average abundance ratio 3.57. (b) C$_2$H$_5$OH versus CH$_3$CHO. (c) $\rm{(CH_2OH)_2}$ versus CH$_2$OHCHO. (d) $\rm{CH_3COCH_3}$ versus CH$_3$OH. (e) $\rm{CH_3OCH_3}$ versus CH$_3$OH, with average abundance ratio 0.12. (f) $\rm{CH_3OCHO}$ versus CH$_3$OH. (g) H$_2$S versus SO$_2$, with average abundance ratio 0.31. (h) NH$_2$CHO versus HNCO, with average abundance ratio 0.36. (i) $\rm{C_2H_5CN}$ versus $\rm{C_2H_3CN}$, with average abundance ratio of 1.28.}
\label{fig:colden-corr}
\end{figure*}

We use our measured column densities in G28.2 HMC1, HMC2 and HMC3 and in three other HMCs that have similar spatial scales (Orion KL - hot core; G10.47; G328.2) to explore potential correlations and abundance ratios between chemically related species. These results provide constraints for astrochemical models of hot cores.
Figure \ref{fig:colden-corr} shows some of our explored correlations.

The measured column density and abundance of CH$_3$OH, as determined from its line profile, may be subject to potential underestimation due to high line opacity (see Fig. \ref{fig:spectra}). To mitigate this issue, we have measured abundances from the rarer species $^{13}$CH$_3$OH. For this we measure the column density of $^{13}$CH$_3$OH and then use an estimated $^{12}$C/$^{13}$C ratio to derive the column density of CH$_3$OH. The $^{12}$C/$^{13}$C isotopologue ratio is estimated to be 44.2 using the relation of \cite{milam2005} and setting the Galactocentric distance of the source to be 4.11 kpc. The column densities of $\rm{^{13}CH_3OH}$ are measured as $\rm{3\times10^{17}\ cm^{-2}}$ towards HMC1 and $\rm{5\times10^{16}\ cm^{-2}}$ towards HMC2.

A clear correlation is observed for H$_2$CO and CH$_3$OH (Fig. \ref{fig:colden-corr}a). These two species are chemically linked (see \S\ref{sec:3_3_3}) and the CH$_3$OH column density is always higher (on average by a factor of 3.5) compared to its precursor, H$_2$CO. $\rm{C_2H_5OH}$ and CH$_3$CHO also show some correlation, except source G10.47 (Fig. \ref{fig:colden-corr}b). 
$\rm{(CH_2OH)_2}$ and CH$_2$OHCHO show a more scattered relation 
(Fig. \ref{fig:colden-corr}c). $\rm{CH_3OCH_3}$ and CH$_3$OH show a clear correlation (with the latter more abundant by an average factor of  8.3), which could be due to the fact that they are produced efficiently from the same precursor, CH$_3$O. The peptide bond containing species HNCO and NH$_2$CHO are directly linked via their formation pathways (see \S\ref{3_3_13}) and we see a clear linear correlation between them, such that the abundance of NH$_2$CHO is on average about 0.36 times that of HNCO (see Fig. \ref{fig:colden-corr}h). 
Finally, a near one-to-one relation is observed between $\rm{C_2H_5CN}$ and $\rm{C_2H_3CN}$, which are precursors of branched-chain molecules \citep{bell17}, with the abundance of the former about 1.3 times the latter on average.

\subsection{Comparison between N-bearing and O-bearing COMs}

Both O- and N-bearing species are well-known tracers of hot molecular HMCs \citep[e.g.,][]{black87,bell14,bell17,rivi17,gora21}. \cite{black87} observed various N- and O-bearing molecules in Orion KL with different radial velocities and found N-bearing species are enhanced in the Orion Hot Core and O-bearing molecules in the compact ridge. Later, astrochemical modeling results showed that such chemical differences could arise due to different physical conditions and/or evolutionary stage \citep[e.g.,][]{charn92,case93}.
To investigate whether the observed chemical differentiation between O- and N-bearing molecules is a general property of HMC or not, \cite{font07} studied N and O-bearing molecules in several HMCs. Similar to our results of \S\ref{sec:column}, they found $\rm{C_2H_3CN}$ and $\rm{C_2H_5CN}$ are strongly correlated. Later, results of interferometric observations suggested O-bearing species are enhanced in the Orion Compact Ridge and N-bearing species are rich in the Orion Hot Core \citep[e.g.,][]{frid08,peng13,feng15}, which further supported the previous results of \citet{black87}. \citep{suzu18} studied various N- and O-bearing species in several HMCs using NRO 45m single dish data. They did not find any correlation between N and O-bearing species, instead finding a correlation between N-bearing molecules with other N-bearing molecules. 

In G28.2 we find correlations of the spatial distributions between some of the O and N-bearing molecules. Emission from all the detected high excitation transitions of O-bearing and N-bearing COMs is compact and bright toward HMC1 and HMC2, except for $\rm{C_2H_3CN}$. 
This suggests that O- and N-bearing COMs show similar morphological structures and are bright toward HMC1, indicating the active sites of formation of both types of species. In conclusion, O- and N-bearing molecules show similar distribution hints of similar chemical origins i.e., formation on the grains followed by sublimation into the gas phase. However, it is possible that higher angular resolution observations with the sensitivity to detect the species will reveal spatial variations on scales smaller than $\sim 0.2\arcsec$, i.e., $\sim 1,000\:$au.

\subsection{Shock-induced chemistry \label{shock}}

\begin{figure*}
\includegraphics[width=\textwidth]{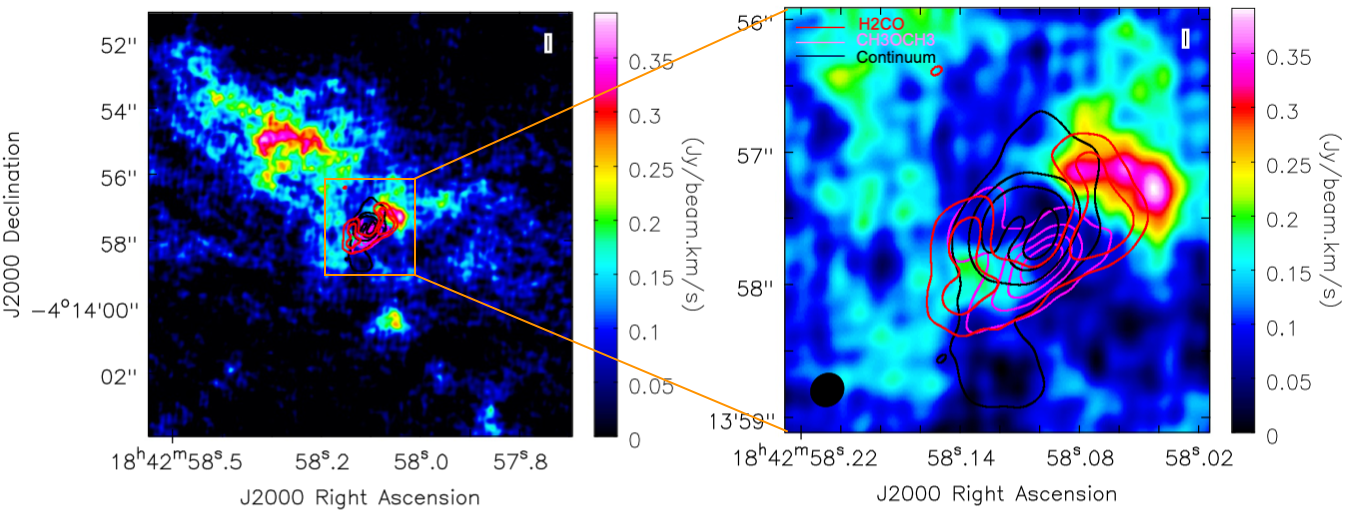}
\caption{The color image represents the SiO emission, which mainly traces shocks. Red and magenta contour represents high excitation H$_2$CO and CH$_3$OCH$_3$, respectively. The black contour represents the continuum contours.}
\label{fig:shock}
\end{figure*}

In addition to the emission from HMC1, HMC2 and HMC3 in and around the continuum ring, there is also some emission from molecular gas, including COMs, from more extended regions, especially to the NE in the direction of the known blueshifted outflow. In the inner regions of this source, SiO emission is primarily tracing outflow activity. Interestingly, as shown in Figure~\ref{fig:shock}, SiO is also seen at the positions of the all the HMCs, in particular with local peaks seen towards HMC1 and HMC3.
Therefore, in addition to the energy from radiative heating, the molecular gas around the protostar may also be receiving significant heating from outflow-driven shocks.

It is well known that $\rm{H_2CO}$ and $\rm{CH_3OH}$ are mainly produced in grain mantles in the ice phase and then later evaporated to the gas phase. In addition to emission from hot core regions, these two species have also been observed in several outflows, with abundance enhanced by factor of $\sim$100 \citep[e.g.,][]{bach01,garay02,jorg04, maret05} compared to the hot core region.
In particular, low excitation transitions of $\rm{H_2CO}$ and $\rm{CH_3OH}$ have been found to be good tracers of outflow shocks \citep[e.g.,][]{arce07,tani20,zhang20}.

In G28.2, we find some morphological similarity of the low excitation $\rm{H_2CO}$ and$\rm{CH_3OH}$ transitions with the SiO emission.
Thus it is evident that in the extended regions, the main reason for the enhancement of $\rm{H_2CO}$ and$\rm{CH_3OH}$ is due to the presence of shocks. 
On the other hand, we also see extended emission of H$_2$S, whose enhanced emission peak is overlapped with the SiO emission peak (see \S\ref{sec:sulfur} for further discussion).

\subsection{Sulfur-bearing species\label{sec:sulfur}}

In diffuse clouds, observed abundances of sulfur reflect the elemental abundance ($\sim$10$^{-5}$) \citep[e.g.,][]{sava96,howk06}. However, the main reservoir of sulfur is still puzzling \citep[e.g.,][]{tief94,wood15}. In dense molecular clouds, S-bearing molecules are thought to be depleted on to grain surfaces, which could be the main reason why the observed gas-phase abundance of sulfur does not reflect its cosmic elemental abundance. It remains a long-standing open question as to which species are the main S-bearing species in these environments.
So far, only two S-bearing species have been identified on grain surfaces: OCS and SO$_2$ \citep{palu95,palu97,boog97}. However, 
the abundance of SO$_2$ is not thought to be make the major contribution to the S reservoir.
Instead, based on various observational and especially modeling results, H$_2$S and OCS have been thought to be the most important S=beraing species.
\citep[e.g.,][]{char97,wake04,podi14}.

Previously, two S-bearing molecules, SO$_2$ ($J'_{K'_aK'_c}-J''_{K''_aK''_c}$ = 11,1,11 - 10,0,10)
and OCS ($J'_{K'_aK'_c}-J''_{K''_aK''_c}$ = 19 - 18)
were detected toward G28.2 \citep{qin08,klaa09}. \cite{klaa09} analyzed sub-millimeter array (SMA) data and studied moment maps of SO$_2$ and OCS. They found different distributions of emissions for these two species (see their Fig. 2). They concluded that these molecules trace gas surrounding the HII region that appears to be rotating. They also noticed different velocity gradients for these two species. However, these features were only marginally resolved by their observations (i.e., beam of 1$\arcsec\sim5700\:$au). Our observations have $\sim$5 times higher angular resolution (i.e., beam of 0.22\arcsec $\sim$ 1250 au). We have detected and spatially resolved three S-bearing molecules: $^{18}$OCS; SO$_2$; and H$_2$S (see Table \ref{table:LINES}). Among these, H$_2$S and $^{18}$OCS are first detections in this source, while the high excitation transition of SO$_2$ is also a new detection. These three S-bearing species show very different spatial distributions, indicating they trace different physical environments and that they most likely have different chemical origins. 

Figure \ref{fig:moment0-obearing}d shows SO$_2$ ($E_{\rm up}=248\:$K) emission, which traces the mm continuum ring in the inner regions of the protostellar core. Figure~\ref{figh:moment1}b shows that this gas has complex kinematics, with the velocity centroid varying by several km/s at different locations around the ring.

Our $^{18}$OCS emission shows a similar distribution to OCS \citep[see Fig. 2,][]{klaa09}, which itself is similar to other COMs seen in this source such as methanol, acetaldehyde and dimethyl ether. Similar distributions of these species may indicate a similar origin, i.e., on grain surfaces and then sublimated to the gas phase. H$_2$S and OCS are thought to be the major reservoir of sulfur, which might have been locked onto the grain surface \citep{char97,wake04,podi14}. We see a completely different distribution of H$_2$S (see Fig.~\ref{fig:moment0-simple}d) as compared to OCS and other COMs, indicating different formation paths, i.e., $\rm H_2S$ may have been produced in the gas phase rather than on grain surfaces. Some morphological similarities between emission peaks of SiO and H$_2$S suggests that H$_2$S also traces shocks in the outflowing gas.

\section{Conclusions \label{sec:conclusions}}

We have analyzed high angular resolution ($\sim0.2\arcsec)$ ALMA 1.3~mm (Band 6) data of G28.2, an isolated massive protostar, to study the astrochemical inventory and characterize conditions in the dense molecular gas that is forming the star. The main results are as follows.

\begin{itemize} 

\item{Based on an investigation of spectra and emission maps of different observed species, we identify three main hot molecular cores, HMC1, HMC2 and HMC3, in the immediate vicinity (within $\sim 3,000\:$au in projection) of the massive protostar. HMC1 and HMC2 are part of the previously identified mm continuum ring-like structure, while HMC3 is slightly further away. These cores are estimated to have temperatures of 300~K, 250~K and 70~K, respectively, have mass surface densities of $\sim 10\:{\rm g\:cm}^{-2}$, densities $n_{\rm H}\sim 10^9\:{\rm cm}^{-3}$ and masses from $\sim 1$ to $\sim 4\:M_\odot$. The total mass of molecular gas within 3,000~au is estimated to be $\sim 25\:M_\odot$, which is similar to that of the central protostar ($m_*\sim 40\:M_\odot$) and this provides an important constraint on theoretical models of massive star formation.}

\item{There is evidence for systematic increase in the offset of line emission from the main continuum peak with decreasing upper state energy of the transition. This is tentative evidence of temperature gradients in the cores, indicating that they are heated externally by the central protostar. The presence of outflow tracers (SiO and high velocity CO) in the region also indicates a possibility of shock heating in and around the cores. Our high spatial resolution also allows us to see the detailed morphologies of high excitation species and larger COMs concetrated in the inner region, in and around the mm continuum ring, and the presence of lower excitation emission from relatively simple species in the more extended, $\sim 0.1\:$pc scale core envelope. There is also evidence for larger scale outflow shocks enhancing certain species, including H$_2$S and low excitation transitions of $\rm{H_2CO}$ and $\rm{CH_3OH}$, in the extended region.}

\item{HMC1 is the most chemically rich, followed by HMC2 and then HMC3. Overall 8 simple species and 14 COMs have been identified, making this source relatively chemically rich compared to other massive protostars that have been observed with the same type of ALMA observation.}

\item {Assuming LTE and optically thin conditions, we have estimate the column densities of all the observed species by comparison with synthetic spectra. We have thus estimated absolute abundances and abundance ratios that provide useful constraints for astrochemical models of hot cores. Including data for an additional three other HMCs from the literature, we find evidence for correlations between column densities of certain pairs of species, i.e., approximately constant abundance ratios, which is evidence for their formation in linked astrochemical processes. These pairs include: CH$_3$OH versus H$_2$CO, with average abundance ratio 3.57; $\rm{CH_3OCH_3}$ versus CH$_3$OH, with average abundance ratio 0.12; H$_2$S versus SO$_2$, with average abundance ratio 0.31; NH$_2$CHO versus HNCO, with average abundance ratio 0.36; and $\rm{C_2H_5CN}$ versus $\rm{C_2H_3CN}$, with average abundance ratio of 1.28.}

\item{Various sulfur-bearing exhibit a variety of morphological structures in G28.2. OCS has a morphology concentrated toward the central mm continuum ring, including HMC1 and HMC2, that is similar to many of the COM species, indicating an origin in the evaporation of grain mantles.
SO$_2$, seen via a single high excitation transition, is also concentrated in and around the mm continuum ring.
However, H$_2$S shows much more extended emission, similar to $\rm{H_2CO}$ and $\rm{CH_3OH}$, on scales of $\sim0.1\:$pc and likely influenced by protostellar outflow shocks.} 

\end{itemize}

\acknowledgments
This paper makes use of the following ALMA data: ADS/JAO.ALMA\#2015.1.01454.S, ADS/JAO.ALMA\#2016.1.00125.S. ALMA is a partnership of ESO (representing its member states), NSF (USA) and NINS (Japan), together with NRC (Canada), MOST and ASIAA (Taiwan), and KASI (Republic of Korea), in cooperation with the Republic of Chile. The Joint ALMA Observatory is operated by ESO, AUI/NRAO and NAOJ. PG and GC acknowledges support from a Chalmers Initiative on Cosmic Origins (CICO) postdoctoral fellowship. JCT acknowledges support from ERC Advanced Grant MSTAR, VR grant Fire from Ice, and NSF grant AST-2206450. We thank Jan Bredeh\"oft for helpful discussions. YZ acknowledges the sponsorship from Yangyang Development Fund.  R.F. acknowledges support from the grants Juan de la Cierva FJC2021-046802-I, PID2020-114461GB-I00 and CEX2021-001131- S funded by MCIN/AEI/10.13039/501100011033 and by “European Union NextGenerationEU/PRTR” and grant P20-00880 from the Consejería de Transformación Económica, Industria, Conocimiento y Universidades of the Junta de Andalucía. R.F. also acknowledges funding from the European Union’s Horizon 2020 research and innovation programme under the Marie Sklodowska-Curie grant agreement No 101032092.

\bibliographystyle{aasjournal}
\bibliography{References}{}

\begin{thebibliography}{}
\expandafter\ifx\csname natexlab\endcsname\relax\def\natexlab#1{#1}\fi
\providecommand{\url}[1]{\href{#1}{#1}}
\providecommand{\dodoi}[1]{doi:~\href{http://doi.org/#1}{\nolinkurl{#1}}}
\providecommand{\doeprint}[1]{\href{http://ascl.net/#1}{\nolinkurl{http://ascl.net/#1}}}
\providecommand{\doarXiv}[1]{\href{https://arxiv.org/abs/#1}{\nolinkurl{https://arxiv.org/abs/#1}}}

\bibitem[{{Arce} {et~al.}(2007){Arce}, {Shepherd}, {Gueth}, {Lee}, {Bachiller},
  {Rosen}, \& {Beuther}}]{arce07}
{Arce}, H.~G., {Shepherd}, D., {Gueth}, F., {et~al.} 2007, in Protostars and
  Planets V, ed. B.~{Reipurth}, D.~{Jewitt}, \& K.~{Keil}, 245.
\newblock \doarXiv{astro-ph/0603071}

\bibitem[{{Bachiller} {et~al.}(2001){Bachiller}, {P{\'e}rez Guti{\'e}rrez},
  {Kumar}, \& {Tafalla}}]{bach01}
{Bachiller}, R., {P{\'e}rez Guti{\'e}rrez}, M., {Kumar}, M.~S.~N., \&
  {Tafalla}, M. 2001, \aap, 372, 899, \dodoi{10.1051/0004-6361:20010519}

\bibitem[{{Bally} {et~al.}(2020){Bally}, {Ginsburg}, {Forbrich}, \&
  {Vargas-Gonz{\'a}lez}}]{2020ApJ...889..178B}
{Bally}, J., {Ginsburg}, A., {Forbrich}, J., \& {Vargas-Gonz{\'a}lez}, J. 2020,
  \apj, 889, 178, \dodoi{10.3847/1538-4357/ab65f2}

\bibitem[{{Balucani} {et~al.}(2015){Balucani}, {Ceccarelli}, \&
  {Taquet}}]{balu15}
{Balucani}, N., {Ceccarelli}, C., \& {Taquet}, V. 2015, \mnras, 449, L16,
  \dodoi{10.1093/mnrasl/slv009}

\bibitem[{{Belloche} {et~al.}(2014){Belloche}, {Garrod}, {M{\"u}ller}, \&
  {Menten}}]{bell14}
{Belloche}, A., {Garrod}, R.~T., {M{\"u}ller}, H. S.~P., \& {Menten}, K.~M.
  2014, Science, 345, 1584, \dodoi{10.1126/science.1256678}

\bibitem[{{Belloche} {et~al.}(2009){Belloche}, {Garrod}, {M{\"u}ller},
  {Menten}, {Comito}, \& {Schilke}}]{bell09}
{Belloche}, A., {Garrod}, R.~T., {M{\"u}ller}, H.~S.~P., {et~al.} 2009, \aap,
  499, 215, \dodoi{10.1051/0004-6361/200811550}

\bibitem[{{Belloche} {et~al.}(2019){Belloche}, {Garrod}, {M{\"u}ller},
  {Menten}, {Medvedev}, {Thomas}, \& {Kisiel}}]{bell19}
---. 2019, \aap, 628, A10, \dodoi{10.1051/0004-6361/201935428}

\bibitem[{{Belloche} {et~al.}(2017){Belloche}, {Meshcheryakov}, {Garrod},
  {Ilyushin}, {Alekseev}, {Motiyenko}, {Margul{\`e}s}, {M{\"u}ller}, \&
  {Menten}}]{bell17}
{Belloche}, A., {Meshcheryakov}, A.~A., {Garrod}, R.~T., {et~al.} 2017, \aap,
  601, A49, \dodoi{10.1051/0004-6361/201629724}

\bibitem[{{Blake} {et~al.}(1987){Blake}, {Sutton}, {Masson}, \&
  {Phillips}}]{black87}
{Blake}, G.~A., {Sutton}, E.~C., {Masson}, C.~R., \& {Phillips}, T.~G. 1987,
  \apj, 315, 621, \dodoi{10.1086/165165}

\bibitem[{Bonnell {et~al.}(1998)Bonnell, Bate, \& Zinnecker}]{bonn98}
Bonnell, I.~A., Bate, M.~R., \& Zinnecker, H. 1998, Monthly Notices of the
  Royal Astronomical Society, 298, 93, \dodoi{10.1046/j.1365-8711.1998.01590.x}

\bibitem[{{Bonnell} {et~al.}(2001){Bonnell}, {Clarke}, {Bate}, \&
  {Pringle}}]{bonn01}
{Bonnell}, I.~A., {Clarke}, C.~J., {Bate}, M.~R., \& {Pringle}, J.~E. 2001,
  \mnras, 324, 573, \dodoi{10.1046/j.1365-8711.2001.04311.x}

\bibitem[{{Boogert} {et~al.}(1997){Boogert}, {Schutte}, {Helmich}, {Tielens},
  \& {Wooden}}]{boog97}
{Boogert}, A.~C.~A., {Schutte}, W.~A., {Helmich}, F.~P., {Tielens},
  A.~G.~G.~M., \& {Wooden}, D.~H. 1997, \aap, 317, 929

\bibitem[{{Bouscasse} {et~al.}(2022){Bouscasse}, {Csengeri}, {Belloche},
  {Wyrowski}, {Bontemps}, {G{\"u}sten}, \& {Menten}}]{bous22}
{Bouscasse}, L., {Csengeri}, T., {Belloche}, A., {et~al.} 2022, \aap, 662, A32,
  \dodoi{10.1051/0004-6361/202140519}

\bibitem[{{Bredeh{\"o}ft} {et~al.}(2017){Bredeh{\"o}ft}, {B{\"o}hler},
  {Schmidt}, {Borrmann}, \& {Swiderek}}]{2017ESC.....1...50B}
{Bredeh{\"o}ft}, J.~H., {B{\"o}hler}, E., {Schmidt}, F., {Borrmann}, T., \&
  {Swiderek}, P. 2017, ACS Earth and Space Chemistry, 1, 50,
  \dodoi{10.1021/acsearthspacechem.6b00011}

\bibitem[{{Brouillet} {et~al.}(2015){Brouillet}, {Despois}, {Lu}, {Baudry},
  {Cernicharo}, {Bockel{\'e}e-Morvan}, {Crovisier}, \& {Biver}}]{brou15}
{Brouillet}, N., {Despois}, D., {Lu}, X.~H., {et~al.} 2015, \aap, 576, A129,
  \dodoi{10.1051/0004-6361/201424588}

\bibitem[{{Caselli} {et~al.}(1993){Caselli}, {Hasegawa}, \& {Herbst}}]{case93}
{Caselli}, P., {Hasegawa}, T.~I., \& {Herbst}, E. 1993, \apj, 408, 548,
  \dodoi{10.1086/172612}

\bibitem[{{Cernicharo} {et~al.}(2016){Cernicharo}, {Kisiel}, {Tercero},
  {Kolesnikov{\'a}}, {Medvedev}, {L{\'o}pez}, {Fortman}, {Winnewisser}, {de
  Lucia}, {Alonso}, \& {Guillemin}}]{cern16}
{Cernicharo}, J., {Kisiel}, Z., {Tercero}, B., {et~al.} 2016, \aap, 587, L4,
  \dodoi{10.1051/0004-6361/201527531}

\bibitem[{{Cesaroni} {et~al.}(2010){Cesaroni}, {Hofner}, {Araya}, \&
  {Kurtz}}]{cesa10}
{Cesaroni}, R., {Hofner}, P., {Araya}, E., \& {Kurtz}, S. 2010, \aap, 509, A50,
  \dodoi{10.1051/0004-6361/200912877}

\bibitem[{{Charnley}(1997)}]{char97}
{Charnley}, S.~B. 1997, \apj, 481, 396, \dodoi{10.1086/304011}

\bibitem[{{Charnley} {et~al.}(1992){Charnley}, {Tielens}, \&
  {Millar}}]{charn92}
{Charnley}, S.~B., {Tielens}, A.~G.~G.~M., \& {Millar}, T.~J. 1992, \apjl, 399,
  L71, \dodoi{10.1086/186609}

\bibitem[{{Colzi} {et~al.}(2021){Colzi}, {Rivilla}, {Beltr{\'a}n},
  {Jim{\'e}nez-Serra}, {Mininni}, {Melosso}, {Cesaroni}, {Fontani},
  {Lorenzani}, {S{\'a}nchez-Monge}, {Viti}, {Schilke}, {Testi}, {Alonso}, \&
  {Kolesnikov{\'a}}}]{colzi21}
{Colzi}, L., {Rivilla}, V.~M., {Beltr{\'a}n}, M.~T., {et~al.} 2021, \aap, 653,
  A129, \dodoi{10.1051/0004-6361/202141573}

\bibitem[{{Combes} {et~al.}(1987){Combes}, {Gerin}, {Wootten}, {Wlodarczak},
  {Clausset}, \& {Encrenaz}}]{comb87}
{Combes}, F., {Gerin}, M., {Wootten}, A., {et~al.} 1987, \aap, 180, L13

\bibitem[{{Csengeri} {et~al.}(2019){Csengeri}, {Belloche}, {Bontemps},
  {Wyrowski}, {Menten}, \& {Bouscasse}}]{csen19}
{Csengeri}, T., {Belloche}, A., {Bontemps}, S., {et~al.} 2019, \aap, 632, A57,
  \dodoi{10.1051/0004-6361/201935226}

\bibitem[{{de la Fuente} {et~al.}(2020){de la Fuente}, {Porras}, {Trinidad},
  {Kurtz}, {Kemp}, {Tafoya}, {Franco}, \& {Rodr{\'\i}guez-Rico}}]{fuen20}
{de la Fuente}, E., {Porras}, A., {Trinidad}, M.~A., {et~al.} 2020, \mnras,
  492, 895, \dodoi{10.1093/mnras/stz3482}

\bibitem[{{Enrique-Romero} {et~al.}(2020){Enrique-Romero},
  {{\'A}lvarez-Barcia}, {Kolb}, {Rimola}, {Ceccarelli}, {Balucani}, {Meisner},
  {Ugliengo}, {Lamberts}, \& {K{\"a}stner}}]{rome20}
{Enrique-Romero}, J., {{\'A}lvarez-Barcia}, S., {Kolb}, F.~J., {et~al.} 2020,
  \mnras, 493, 2523, \dodoi{10.1093/mnras/staa484}

\bibitem[{{Feng} {et~al.}(2015){Feng}, {Beuther}, {Henning}, {Semenov},
  {Palau}, \& {Mills}}]{feng15}
{Feng}, S., {Beuther}, H., {Henning}, T., {et~al.} 2015, \aap, 581, A71,
  \dodoi{10.1051/0004-6361/201322725}

\bibitem[{{Fontani} {et~al.}(2007){Fontani}, {Pascucci}, {Caselli}, {Wyrowski},
  {Cesaroni}, \& {Walmsley}}]{font07}
{Fontani}, F., {Pascucci}, I., {Caselli}, P., {et~al.} 2007, \aap, 470, 639,
  \dodoi{10.1051/0004-6361:20077485}

\bibitem[{{Friedel} \& {Snyder}(2008)}]{frid08}
{Friedel}, D.~N., \& {Snyder}, L.~E. 2008, \apj, 672, 962,
  \dodoi{10.1086/523896}

\bibitem[{{Garay} {et~al.}(2002){Garay}, {Mardones}, {Rodr{\'\i}guez},
  {Caselli}, \& {Bourke}}]{garay02}
{Garay}, G., {Mardones}, D., {Rodr{\'\i}guez}, L.~F., {Caselli}, P., \&
  {Bourke}, T.~L. 2002, \apj, 567, 980, \dodoi{10.1086/338668}

\bibitem[{{Garrod}(2013)}]{garr13}
{Garrod}, R.~T. 2013, \apj, 765, 60, \dodoi{10.1088/0004-637X/765/1/60}

\bibitem[{{Garrod} {et~al.}(2017){Garrod}, {Belloche}, {M{\"u}ller}, \&
  {Menten}}]{garr17}
{Garrod}, R.~T., {Belloche}, A., {M{\"u}ller}, H.~S.~P., \& {Menten}, K.~M.
  2017, \aap, 601, A48, \dodoi{10.1051/0004-6361/201630254}

\bibitem[{{Garrod} \& {Herbst}(2006)}]{garr06}
{Garrod}, R.~T., \& {Herbst}, E. 2006, \aap, 457, 927,
  \dodoi{10.1051/0004-6361:20065560}

\bibitem[{{Garrod} {et~al.}(2008){Garrod}, {Widicus Weaver}, \&
  {Herbst}}]{garr08}
{Garrod}, R.~T., {Widicus Weaver}, S.~L., \& {Herbst}, E. 2008, \apj, 682, 283,
  \dodoi{10.1086/588035}

\bibitem[{{Gorai} {et~al.}(2020){Gorai}, {Bhat}, {Sil}, {Mondal}, {Ghosh},
  {Chakrabarti}, \& {Das}}]{gora20}
{Gorai}, P., {Bhat}, B., {Sil}, M., {et~al.} 2020, \apj, 895, 86,
  \dodoi{10.3847/1538-4357/ab8871}

\bibitem[{{Gorai} {et~al.}(2017a){Gorai}, {Das}, {Das}, {Sivaraman}, {Etim}, \&
  {Chakrabarti}}]{gora17a}
{Gorai}, P., {Das}, A., {Das}, A., {et~al.} 2017a, \apj, 836, 70,
  \dodoi{10.3847/1538-4357/836/1/70}

\bibitem[{{Gorai} {et~al.}(2021){Gorai}, {Das}, {Shimonishi}, {Sahu}, {Mondal},
  {Bhat}, \& {Chakrabarti}}]{gora21}
{Gorai}, P., {Das}, A., {Shimonishi}, T., {et~al.} 2021, \apj, 907, 108,
  \dodoi{10.3847/1538-4357/abc9c4}

\bibitem[{{Grudi{\'c}} {et~al.}(2022){Grudi{\'c}}, {Guszejnov}, {Offner},
  {Rosen}, {Raju}, {Faucher-Gigu{\`e}re}, \& {Hopkins}}]{2022MNRAS.512..216G}
{Grudi{\'c}}, M.~Y., {Guszejnov}, D., {Offner}, S. S.~R., {et~al.} 2022,
  \mnras, 512, 216, \dodoi{10.1093/mnras/stac526}

\bibitem[{{Haupa} {et~al.}(2019){Haupa}, {Tarczay}, \& {Lee}}]{haup19}
{Haupa}, K.~A., {Tarczay}, G., \& {Lee}, Y.-P. 2019, J. Am. Chem. Soc., 141,
  11614

\bibitem[{{Herbst} \& {van Dishoeck}(2009)}]{herb09}
{Herbst}, E., \& {van Dishoeck}, E.~F. 2009, \araa, 47, 427,
  \dodoi{10.1146/annurev-astro-082708-101654}

\bibitem[{{Hern{\'a}ndez-Hern{\'a}ndez}
  {et~al.}(2014){Hern{\'a}ndez-Hern{\'a}ndez}, {Zapata}, {Kurtz}, \&
  {Garay}}]{hern14}
{Hern{\'a}ndez-Hern{\'a}ndez}, V., {Zapata}, L., {Kurtz}, S., \& {Garay}, G.
  2014, \apj, 786, 38, \dodoi{10.1088/0004-637X/786/1/38}

\bibitem[{{Howk} {et~al.}(2006){Howk}, {Sembach}, \& {Savage}}]{howk06}
{Howk}, J.~C., {Sembach}, K.~R., \& {Savage}, B.~D. 2006, \apj, 637, 333,
  \dodoi{10.1086/497352}

\bibitem[{{J{\o}rgensen} {et~al.}(2004){J{\o}rgensen}, {Hogerheijde}, {Blake},
  {van Dishoeck}, {Mundy}, \& {Sch{\"o}ier}}]{jorg04}
{J{\o}rgensen}, J.~K., {Hogerheijde}, M.~R., {Blake}, G.~A., {et~al.} 2004,
  \aap, 415, 1021, \dodoi{10.1051/0004-6361:20034216}

\bibitem[{Jørgensen {et~al.}(2020)Jørgensen, Belloche, \& Garrod}]{jorg20}
Jørgensen, J.~K., Belloche, A., \& Garrod, R.~T. 2020, Annual Review of
  Astronomy and Astrophysics, 58, 727,
  \dodoi{10.1146/annurev-astro-032620-021927}

\bibitem[{Klaassen {et~al.}(2009)Klaassen, Wilson, Keto, \& Zhang}]{klaa09}
Klaassen, P.~D., Wilson, C.~D., Keto, E.~R., \& Zhang, Q. 2009, The
  Astrophysical Journal, 703, 1308, \dodoi{10.1088/0004-637x/703/2/1308}

\bibitem[{{Klaassen} {et~al.}(2011){Klaassen}, {Wilson}, {Keto}, {Zhang},
  {Galv{\'a}n-Madrid}, \& {Liu}}]{klaa11}
{Klaassen}, P.~D., {Wilson}, C.~D., {Keto}, E.~R., {et~al.} 2011, \aap, 530,
  A53, \dodoi{10.1051/0004-6361/201016371}

\bibitem[{{Kurtz} {et~al.}(2000){Kurtz}, {Cesaroni}, {Churchwell}, {Hofner}, \&
  {Walmsley}}]{kurt00}
{Kurtz}, S., {Cesaroni}, R., {Churchwell}, E., {Hofner}, P., \& {Walmsley},
  C.~M. 2000, in Protostars and Planets IV, ed. V.~{Mannings}, A.~P. {Boss}, \&
  S.~S. {Russell}, 299--326

\bibitem[{{Law} {et~al.}(2022){Law}, {Tan}, {Gorai}, {Zhang}, {Fedriani},
  {Tafoya}, {Tanaka}, {Cosentino}, {Yang}, {Mardones}, {Beltr{\'a}n}, \&
  {Garay}}]{law22}
{Law}, C.-Y., {Tan}, J.~C., {Gorai}, P., {et~al.} 2022, arXiv e-prints,
  arXiv:2201.01411.
\newblock \doarXiv{2201.01411}

\bibitem[{{Ligterink} {et~al.}(2017){Ligterink}, {Coutens}, {Kofman},
  {M{\"u}ller}, {Garrod}, {Calcutt}, {Wampfler}, {J{\o}rgensen}, {Linnartz}, \&
  {van Dishoeck}}]{ligt17}
{Ligterink}, N.~F.~W., {Coutens}, A., {Kofman}, V., {et~al.} 2017, \mnras, 469,
  2219, \dodoi{10.1093/mnras/stx890}

\bibitem[{{Luo} {et~al.}(2019){Luo}, {Feng}, {Li}, {Qin}, {Peng}, {Tang},
  {Ren}, \& {Shi}}]{luo19}
{Luo}, G., {Feng}, S., {Li}, D., {et~al.} 2019, \apj, 885, 82,
  \dodoi{10.3847/1538-4357/ab45ef}

\bibitem[{Mangum {et~al.}(2005)Mangum, Wootten, Loren, \& Wadiak}]{mangum2005}
Mangum, J.~G., Wootten, A., Loren, R.~B., \& Wadiak, E.~J. 2005, in Molecular
  Clouds in the Milky Way and External Galaxies: Proceeedings of a Symposium
  Held at the University of Massachussets in Amherst, November 2--4, 1987,
  Springer, 263--264

\bibitem[{{Maret} {et~al.}(2005){Maret}, {Ceccarelli}, {Tielens}, {Caux},
  {Lefloch}, {Faure}, {Castets}, \& {Flower}}]{maret05}
{Maret}, S., {Ceccarelli}, C., {Tielens}, A.~G.~G.~M., {et~al.} 2005, \aap,
  442, 527, \dodoi{10.1051/0004-6361:20052899}

\bibitem[{{McKee} \& {Tan}(2003)}]{mcke03}
{McKee}, C.~F., \& {Tan}, J.~C. 2003, \apj, 585, 850, \dodoi{10.1086/346149}

\bibitem[{{McMullin} {et~al.}(2007){McMullin}, {Waters}, {Schiebel}, {Young},
  \& {Golap}}]{mcmu07}
{McMullin}, J.~P., {Waters}, B., {Schiebel}, D., {Young}, W., \& {Golap}, K.
  2007, in Astronomical Society of the Pacific Conference Series, Vol. 376,
  Astronomical Data Analysis Software and Systems XVI, ed. R.~A. {Shaw},
  F.~{Hill}, \& D.~J. {Bell}, 127

\bibitem[{{Milam} {et~al.}(2005){Milam}, {Savage}, {Brewster}, {Ziurys}, \&
  {Wyckoff}}]{milam2005}
{Milam}, S.~N., {Savage}, C., {Brewster}, M.~A., {Ziurys}, L.~M., \& {Wyckoff},
  S. 2005, \apj, 634, 1126, \dodoi{10.1086/497123}

\bibitem[{Minh(2016b)}]{minh16b}
Minh, Y.~C. 2016b, Journal of Physics: Conference Series, 728, 052007,
  \dodoi{10.1088/1742-6596/728/5/052007}

\bibitem[{{Mondal} {et~al.}(2021){Mondal}, {Gorai}, {Sil}, {Ghosh}, {Etim},
  {Chakrabarti}, {Shimonishi}, {Nakatani}, {Furuya}, {Tan}, \& {Das}}]{mond21}
{Mondal}, S.~K., {Gorai}, P., {Sil}, M., {et~al.} 2021, \apj, 922, 194,
  \dodoi{10.3847/1538-4357/ac1f31}

\bibitem[{{Mues}(2021)}]{Mues21}
{Mues}, M. 2021, Masters Thesis, Bremen University, supervised by J. H.
  Bredeh\"oft \& P. Swiderek

\bibitem[{{M{\"u}ller} {et~al.}(2005){M{\"u}ller}, {Schl{\"o}der}, {Stutzki},
  \& {Winnewisser}}]{mull05}
{M{\"u}ller}, H. S.~P., {Schl{\"o}der}, F., {Stutzki}, J., \& {Winnewisser}, G.
  2005, Journal of Molecular Structure, 742, 215,
  \dodoi{10.1016/j.molstruc.2005.01.027}

\bibitem[{{M{\"u}ller} {et~al.}(2001){M{\"u}ller}, {Thorwirth}, {Roth}, \&
  {Winnewisser}}]{mull01}
{M{\"u}ller}, H.~S.~P., {Thorwirth}, S., {Roth}, D.~A., \& {Winnewisser}, G.
  2001, \aap, 370, L49, \dodoi{10.1051/0004-6361:20010367}

\bibitem[{{Myers} {et~al.}(2013){Myers}, {McKee}, {Cunningham}, {Klein}, \&
  {Krumholz}}]{2013ApJ...766...97M}
{Myers}, A.~T., {McKee}, C.~F., {Cunningham}, A.~J., {Klein}, R.~I., \&
  {Krumholz}, M.~R. 2013, \apj, 766, 97, \dodoi{10.1088/0004-637X/766/2/97}

\bibitem[{{Ossenkopf} \& {Henning}(1994)}]{osse94}
{Ossenkopf}, V., \& {Henning}, T. 1994, \aap, 291, 943

\bibitem[{{Pagani} {et~al.}(2019){Pagani}, {Bergin}, {Goldsmith}, {Melnick},
  {Snell}, \& {Favre}}]{paga19}
{Pagani}, L., {Bergin}, E., {Goldsmith}, P.~F., {et~al.} 2019, \aap, 624, L5,
  \dodoi{10.1051/0004-6361/201935267}

\bibitem[{{Palumbo} {et~al.}(1997){Palumbo}, {Geballe}, \& {Tielens}}]{palu97}
{Palumbo}, M.~E., {Geballe}, T.~R., \& {Tielens}, A.~G.~G.~M. 1997, \apj, 479,
  839, \dodoi{10.1086/303905}

\bibitem[{{Palumbo} {et~al.}(1995){Palumbo}, {Tielens}, \& {Tokunaga}}]{palu95}
{Palumbo}, M.~E., {Tielens}, A.~G.~G.~M., \& {Tokunaga}, A.~T. 1995, \apj, 449,
  674, \dodoi{10.1086/176088}

\bibitem[{{Peng} {et~al.}(2013){Peng}, {Despois}, {Brouillet}, {Baudry},
  {Favre}, {Remijan}, {Wootten}, {Wilson}, {Combes}, \& {Wlodarczak}}]{peng13}
{Peng}, T.~C., {Despois}, D., {Brouillet}, N., {et~al.} 2013, \aap, 554, A78,
  \dodoi{10.1051/0004-6361/201220891}

\bibitem[{{Pickett} {et~al.}(1998){Pickett}, {Poynter}, {Cohen}, {Delitsky},
  {Pearson}, \& {M{\"u}ller}}]{pick98}
{Pickett}, H.~M., {Poynter}, R.~L., {Cohen}, E.~A., {et~al.} 1998, \jqsrt, 60,
  883, \dodoi{10.1016/S0022-4073(98)00091-0}

\bibitem[{{Podio} {et~al.}(2014){Podio}, {Lefloch}, {Ceccarelli}, {Codella}, \&
  {Bachiller}}]{podi14}
{Podio}, L., {Lefloch}, B., {Ceccarelli}, C., {Codella}, C., \& {Bachiller}, R.
  2014, \aap, 565, A64, \dodoi{10.1051/0004-6361/201322928}

\bibitem[{{Qin} {et~al.}(2008){Qin}, {Huang}, {Wu}, {Xue}, \& {Chen}}]{qin08}
{Qin}, S.-L., {Huang}, M., {Wu}, Y., {Xue}, R., \& {Chen}, S. 2008, \apjl, 686,
  L21, \dodoi{10.1086/592785}

\bibitem[{{Rivilla} {et~al.}(2017){Rivilla}, {Beltr{\'a}n}, {Cesaroni},
  {Fontani}, {Codella}, \& {Zhang}}]{rivi17}
{Rivilla}, V.~M., {Beltr{\'a}n}, M.~T., {Cesaroni}, R., {et~al.} 2017, \aap,
  598, A59, \dodoi{10.1051/0004-6361/201628373}

\bibitem[{{Rolffs} {et~al.}(2011){Rolffs}, {Schilke}, {Zhang}, \&
  {Zapata}}]{rolf11}
{Rolffs}, R., {Schilke}, P., {Zhang}, Q., \& {Zapata}, L. 2011, \aap, 536, A33,
  \dodoi{10.1051/0004-6361/201117112}

\bibitem[{{Ruaud} {et~al.}(2015){Ruaud}, {Loison}, {Hickson}, {Gratier},
  {Hersant}, \& {Wakelam}}]{ruau15}
{Ruaud}, M., {Loison}, J.~C., {Hickson}, K.~M., {et~al.} 2015, \mnras, 447,
  4004, \dodoi{10.1093/mnras/stu2709}

\bibitem[{{Savage} \& {Sembach}(1996)}]{sava96}
{Savage}, B.~D., \& {Sembach}, K.~R. 1996, \araa, 34, 279,
  \dodoi{10.1146/annurev.astro.34.1.279}

\bibitem[{{Sewilo} {et~al.}(2004){Sewilo}, {Churchwell}, {Kurtz}, {Goss}, \&
  {Hofner}}]{sewi04}
{Sewilo}, M., {Churchwell}, E., {Kurtz}, S., {Goss}, W.~M., \& {Hofner}, P.
  2004, \apj, 605, 285, \dodoi{10.1086/382268}

\bibitem[{{Sewi{\l}o} {et~al.}(2008){Sewi{\l}o}, {Churchwell}, {Kurtz}, {Goss},
  \& {Hofner}}]{sewi08}
{Sewi{\l}o}, M., {Churchwell}, E., {Kurtz}, S., {Goss}, W.~M., \& {Hofner}, P.
  2008, \apj, 681, 350, \dodoi{10.1086/588422}

\bibitem[{{Singh} {et~al.}(2022){Singh}, {Fabian Kleimeier}, {Eckhardt}, \&
  {Kaiser}}]{sing22}
{Singh}, S.~K., {Fabian Kleimeier}, N., {Eckhardt}, A.~K., \& {Kaiser}, R.~I.
  2022, \apj, 941, 103, \dodoi{10.3847/1538-4357/ac8c92}

\bibitem[{{Skouteris} {et~al.}(2019){Skouteris}, {Balucani}, {Ceccarelli},
  {Faginas Lago}, {Codella}, {Falcinelli}, \& {Rosi}}]{skou19}
{Skouteris}, D., {Balucani}, N., {Ceccarelli}, C., {et~al.} 2019, \mnras, 482,
  3567, \dodoi{10.1093/mnras/sty2903}

\bibitem[{{Sollins} {et~al.}(2005){Sollins}, {Zhang}, {Keto}, \& {Ho}}]{soll05}
{Sollins}, P.~K., {Zhang}, Q., {Keto}, E., \& {Ho}, P. T.~P. 2005, \apj, 631,
  399, \dodoi{10.1086/432503}

\bibitem[{{Suzuki} {et~al.}(2018){Suzuki}, {Ohishi}, {Saito}, {Hirota},
  {Majumdar}, \& {Wakelam}}]{suzu18}
{Suzuki}, T., {Ohishi}, M., {Saito}, M., {et~al.} 2018, \apjs, 237, 3,
  \dodoi{10.3847/1538-4365/aac8db}

\bibitem[{{Tan} {et~al.}(2014){Tan}, {Beltr{\'a}n}, {Caselli}, {Fontani},
  {Fuente}, {Krumholz}, {McKee}, \& {Stolte}}]{2014prpl.conf..149T}
{Tan}, J.~C., {Beltr{\'a}n}, M.~T., {Caselli}, P., {et~al.} 2014, in Protostars
  and Planets VI, ed. H.~{Beuther}, R.~S. {Klessen}, C.~P. {Dullemond}, \&
  T.~{Henning}, 149, \dodoi{10.2458/azu_uapress_9780816531240-ch007}

\bibitem[{{Taniguchi} {et~al.}(2020){Taniguchi}, {Plunkett}, {Herbst},
  {Dobashi}, {Shimoikura}, {Nakamura}, \& {Saito}}]{tani20}
{Taniguchi}, K., {Plunkett}, A., {Herbst}, E., {et~al.} 2020, \mnras, 493,
  2395, \dodoi{10.1093/mnras/staa012}

\bibitem[{{Taquet} {et~al.}(2016){Taquet}, {Wirstr{\"o}m}, \&
  {Charnley}}]{taqu16}
{Taquet}, V., {Wirstr{\"o}m}, E.~S., \& {Charnley}, S.~B. 2016, \apj, 821, 46,
  \dodoi{10.3847/0004-637X/821/1/46}

\bibitem[{{Tieftrunk} {et~al.}(1994){Tieftrunk}, {Pineau des Forets},
  {Schilke}, \& {Walmsley}}]{tief94}
{Tieftrunk}, A., {Pineau des Forets}, G., {Schilke}, P., \& {Walmsley}, C.~M.
  1994, \aap, 289, 579

\bibitem[{{van der Tak}(2004)}]{vand04}
{van der Tak}, F.~F.~S. 2004, in Star Formation at High Angular Resolution, ed.
  M.~G. {Burton}, R.~{Jayawardhana}, \& T.~L. {Bourke}, Vol. 221, 59.
\newblock \doarXiv{astro-ph/0309152}

\bibitem[{{Vastel} {et~al.}(2015){Vastel}, {Bottinelli}, {Caux}, {Glorian}, \&
  {Boiziot}}]{vast15}
{Vastel}, C., {Bottinelli}, S., {Caux}, E., {Glorian}, J.~M., \& {Boiziot}, M.
  2015, in SF2A-2015: Proceedings of the Annual meeting of the French Society
  of Astronomy and Astrophysics, 313--316

\bibitem[{{Vazart} {et~al.}(2020){Vazart}, {Ceccarelli}, {Balucani}, {Bianchi},
  \& {Skouteris}}]{vazart2020}
{Vazart}, F., {Ceccarelli}, C., {Balucani}, N., {Bianchi}, E., \& {Skouteris},
  D. 2020, \mnras, 499, 5547, \dodoi{10.1093/mnras/staa3060}

\bibitem[{{Wakelam} {et~al.}(2004){Wakelam}, {Caselli}, {Ceccarelli}, {Herbst},
  \& {Castets}}]{wake04}
{Wakelam}, V., {Caselli}, P., {Ceccarelli}, C., {Herbst}, E., \& {Castets}, A.
  2004, \aap, 422, 159, \dodoi{10.1051/0004-6361:20047186}

\bibitem[{{Walsh} {et~al.}(2003){Walsh}, {Macdonald}, {Alvey}, {Burton}, \&
  {Lee}}]{wals03}
{Walsh}, A.~J., {Macdonald}, G.~H., {Alvey}, N.~D.~S., {Burton}, M.~G., \&
  {Lee}, J.~K. 2003, \aap, 410, 597, \dodoi{10.1051/0004-6361:20031191}

\bibitem[{{Woods} {et~al.}(2015){Woods}, {Occhiogrosso}, {Viti},
  {Ka{\v{n}}uchov{\'a}}, {Palumbo}, \& {Price}}]{wood15}
{Woods}, P.~M., {Occhiogrosso}, A., {Viti}, S., {et~al.} 2015, \mnras, 450,
  1256, \dodoi{10.1093/mnras/stv652}

\bibitem[{{Zhang} {et~al.}(2021){Zhang}, {Zavagno}, {L{\'o}pez-Sepulcre},
  {Liu}, {Louvet}, {Figueira}, {Russeil}, {Wu}, {Yuan}, \& {Pillai}}]{zhang20}
{Zhang}, S., {Zavagno}, A., {L{\'o}pez-Sepulcre}, A., {et~al.} 2021, \aap, 646,
  A25, \dodoi{10.1051/0004-6361/202038421}

\bibitem[{{Zhang} {et~al.}(2019){Zhang}, {Tan}, {Tanaka}, {De Buizer}, {Liu},
  {Beltr{\'a}n}, {Kratter}, {Mardones}, \& {Garay}}]{2019NatAs...3..517Z}
{Zhang}, Y., {Tan}, J.~C., {Tanaka}, K. E.~I., {et~al.} 2019, Nature Astronomy,
  3, 517, \dodoi{10.1038/s41550-019-0718-y}

\end{thebibliography}

\appendix
\restartappendixnumbering

\section{Moment 1 maps of different species}\label{sect:moment1}

\begin{figure*}
\includegraphics[width=\textwidth]{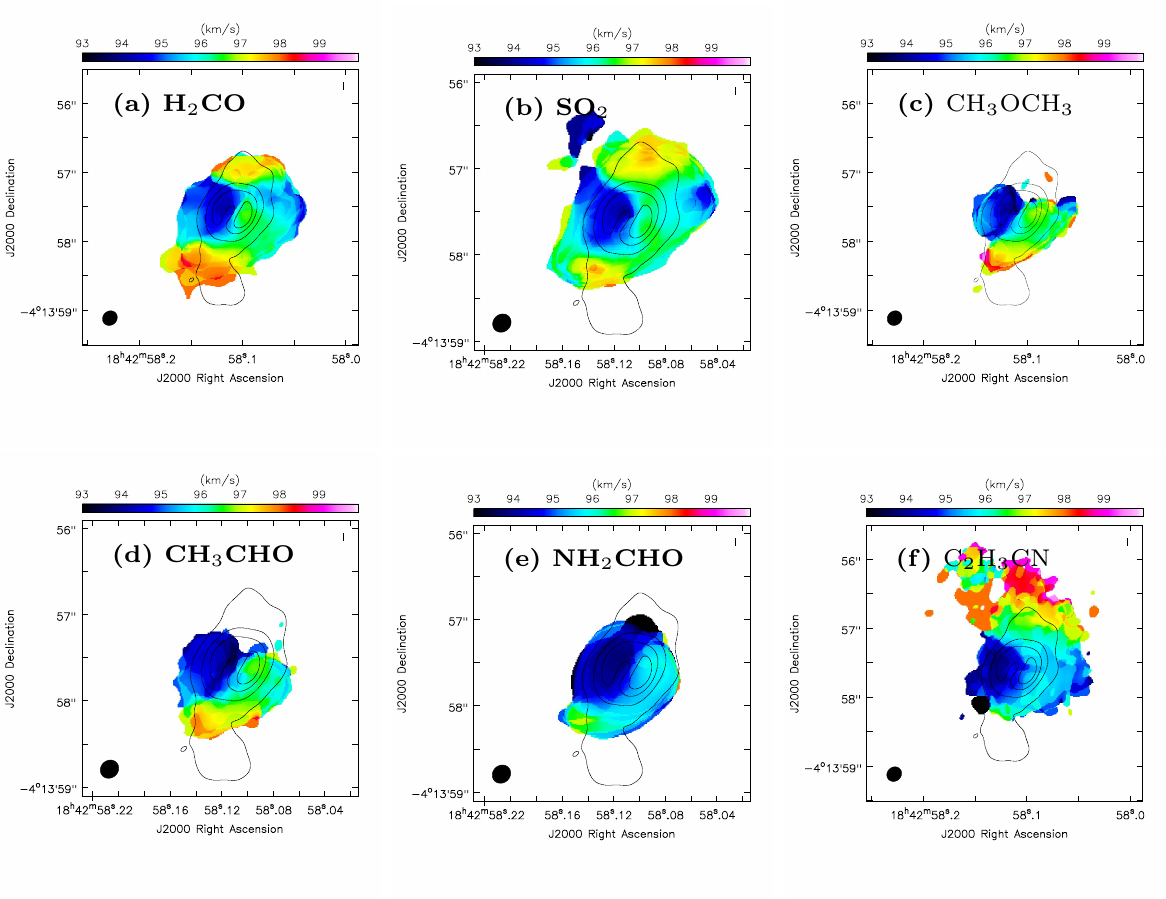}
\caption{Moment 1 maps of: (a) $\rm{H_2CO}$ ($E_{\rm up}=173\:$K); (b) SO$_2$ ($E_{\rm up}=248\:$K); (c) $\rm{CH_3OCH_3}$ ($E_{\rm up}=253\:$K); (d) $\rm{CH_3CHO}$ ($E_{\rm up}=93\:$K); (e) NH$_2$CHO ($E_{\rm up}=60\:$K); and (f) $\rm{C_2H_3CN}$ ($E_{\rm up}=145\:$K).}
\label{figh:moment1}
\end{figure*}

Here, in Figure~\ref{figh:moment1}, we present first moment maps, i.e., showing average velocity, of six of the species identified in G28.2.

\section{LTE model spectra}\label{sect:lte2_abundance}

\begin{figure*}
\centering
\includegraphics[width=17.8cm, height=22cm]{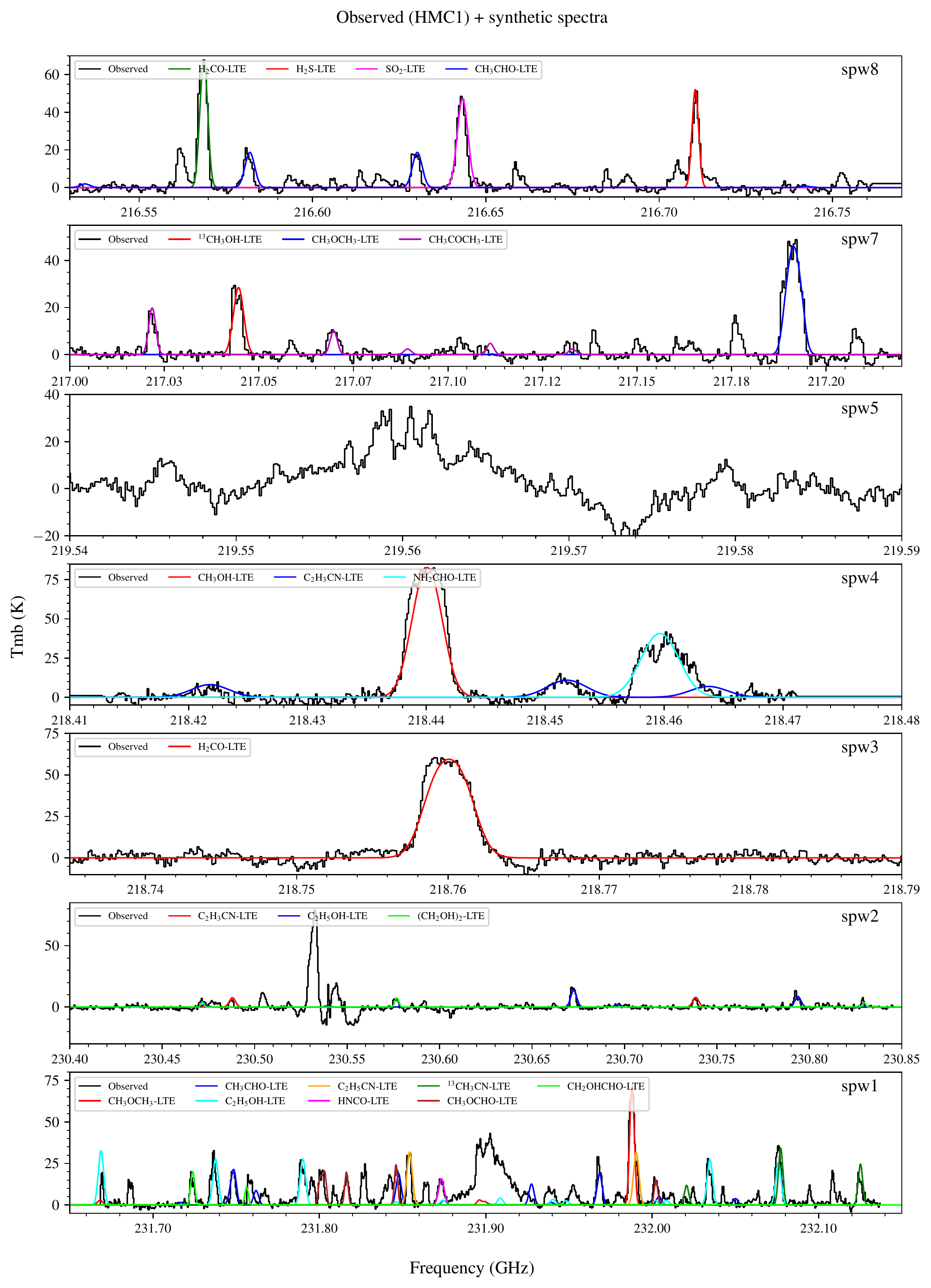}
\caption{Continuum-subtracted spectra towards the G28.2 HMC1. Spectral windows 1, 2, 3, 4, 5, 7, 8 
are shown from bottom to top. The frequencies are in the source rest frame. Synthetic spectra of different species are overlaid on the top of observed spectra, as labelled.}
\label{fig:synthetic1}
\end{figure*}

\begin{figure*}
\centering
\includegraphics[width=17.8cm, height=22cm]{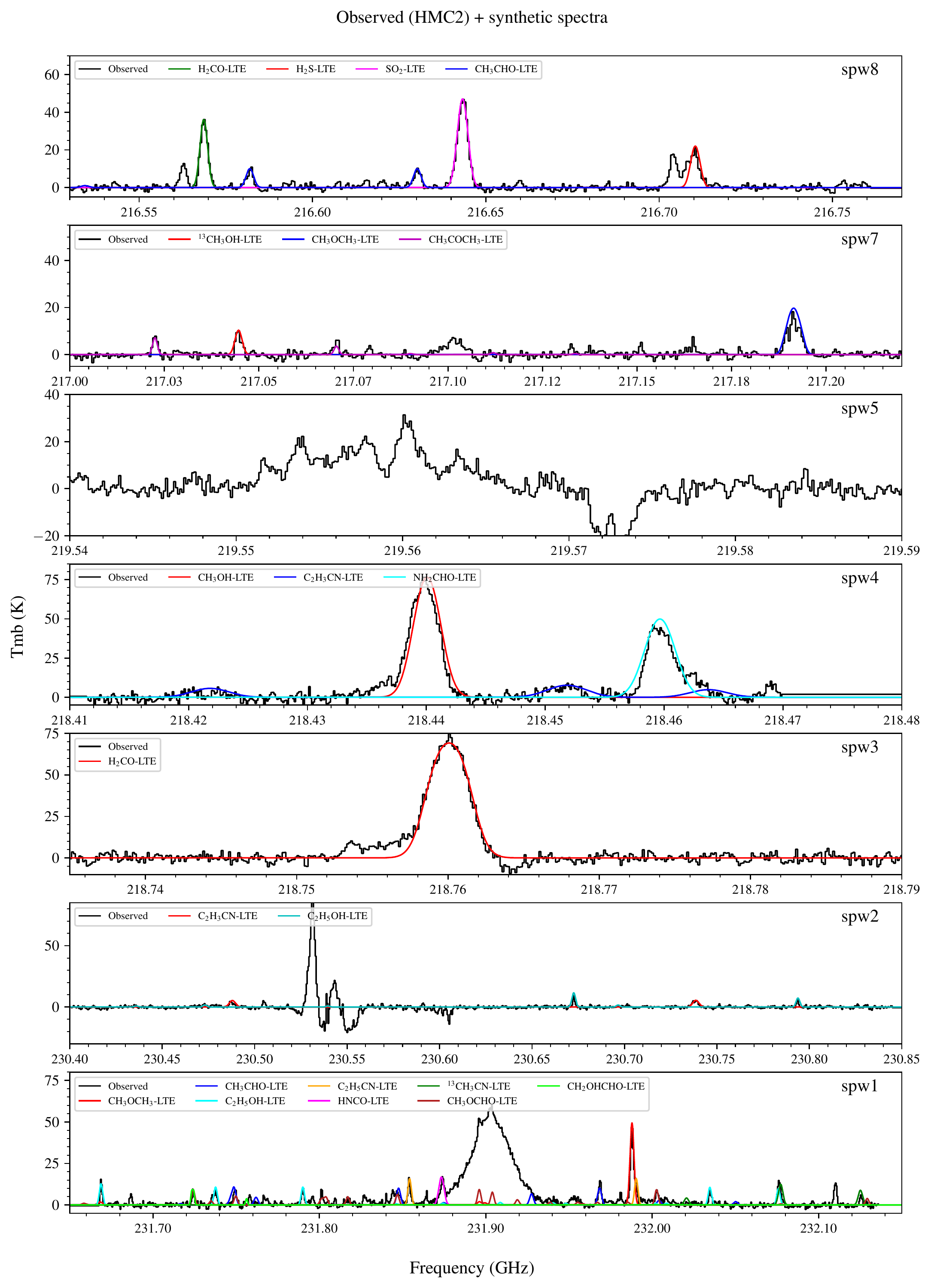}
\caption{Continuum-subtracted spectra towards the G28.2 HMC2. Spectral windows 1, 2, 3, 4, 5, 7, 8 
are shown from bottom to top. The frequencies are in the source rest frame. Synthetic spectra of different species are overlaid on the top of observed spectra, as labelled.}
\label{fig:synthetic2}
\end{figure*}

\begin{figure*}
\centering
\includegraphics[width=17.8cm, height=22cm]{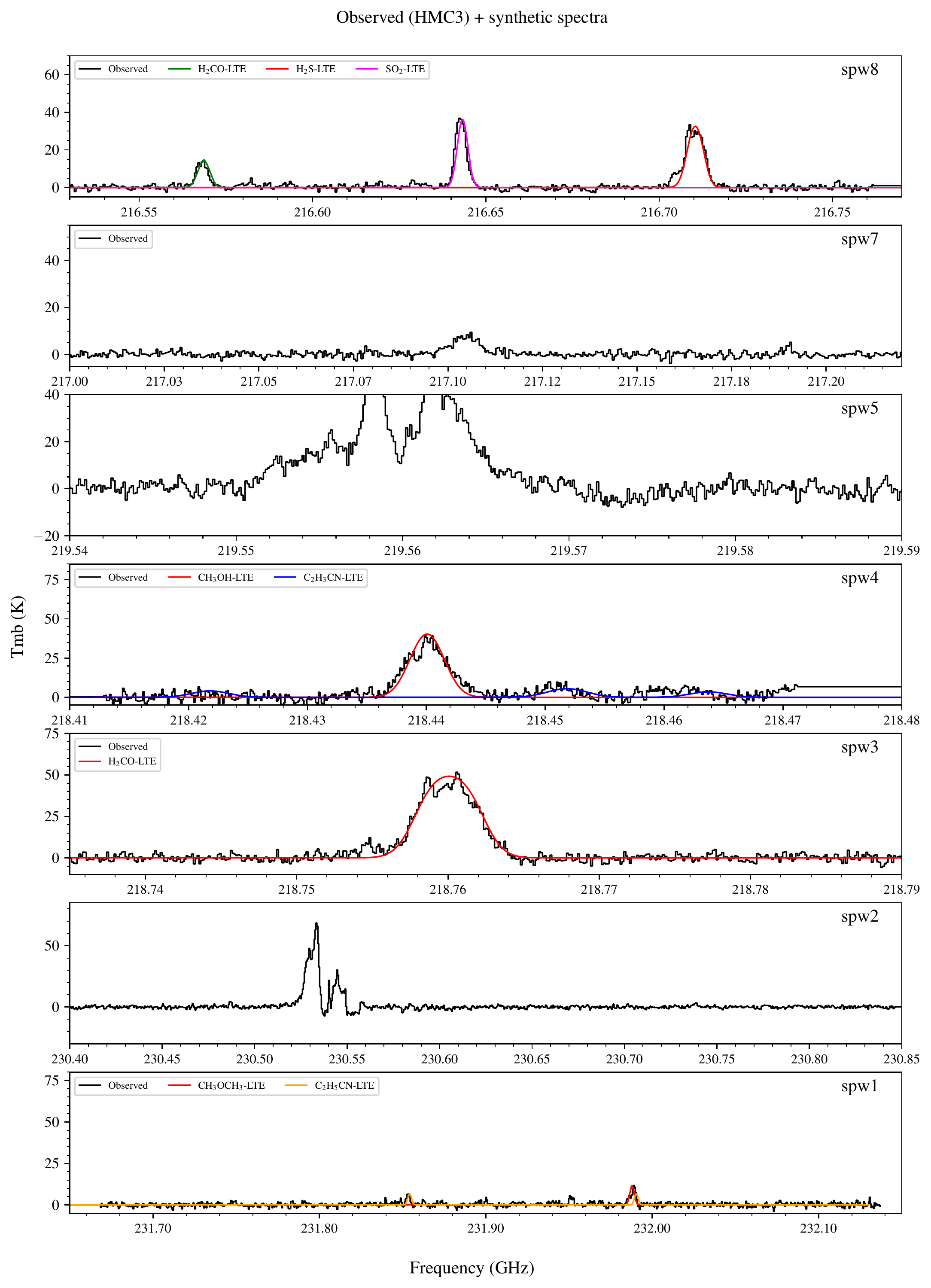}
\caption{Continuum-subtracted spectra towards the G28.2 HMC3. Spectral windows 1, 2, 3, 4, 5, 7, 8 
are shown from bottom to top. The frequencies are in the source rest frame. Synthetic spectra of different species are overlaid on the top of observed spectra, as labelled.}
\label{fig:synthetic3}
\end{figure*}

In Figures \ref{fig:synthetic1}, \ref{fig:synthetic2} and \ref{fig:synthetic3} we show observed spectra, together with synthetic model LTE spectra for HMC1, HMC2 and HMC3, respectively. 

\end{document}